\documentstyle[psfig,harvard]{mn}

\title{The Nature of Galaxy Bias and Clustering}
\author[A.~J.~Benson, S.~Cole, C.~S.~Frenk, C.~M.~Baugh, and
C.~G.~Lacey]{A.~J.~Benson$^{1,2}$, S.~Cole$^{1,3}$, C.~S.~Frenk$^{1,4}$, C.~M.~Baugh$^{1,5}$, and C.~G.~Lacey$^{1,6,7}$ \\
1. Physics Department, University of Durham, Durham DH1 3LE, UK. \\
2. E-mail: A.J.Benson@durham.ac.uk \\
3. E-mail: Shaun.Cole@durham.ac.uk \\
4. E-mail: C.S.Frenk@durham.ac.uk \\
5. E-mail: C.M.Baugh@durham.ac.uk \\
6. Theoretical Astrophysics Center, Copenhagen, Denmark \\
7. E-mail: C.G.Lacey@durham.ac.uk \\
}

\begin{document}

\maketitle

\begin{abstract}

We have used a combination of high resolution cosmological N-body
simulations and semi-analytic modelling of galaxy formation to
investigate the processes that determine the spatial distribution of
galaxies in cold dark matter (CDM) models and its relation to the
spatial distribution of dark matter. The galaxy distribution depends
sensitively on the efficiency with which galaxies form in halos of
different mass. In small mass halos, galaxy formation is inhibited by
the reheating of cooled gas by feedback processes, whereas in large
mass halos, it is inhibited by the long cooling time of the gas. As a
result, the mass-to-light ratio of halos has a deep minimum at the
halo mass, $\sim 10^{12} M_{\odot}$, associated with $L_*$ galaxies,
where galaxy formation is most efficient. This dependence of
galaxy formation efficiency on halo mass leads to a scale-dependent
bias in the distribution of galaxies relative to the distribution of
mass. On large scales, the bias in the galaxy distribution is related
in a simple way to the bias in the distribution of massive halos.  On
small scales, the correlation function is determined by the interplay
between various effects including the spatial exclusion of dark matter
halos, the distribution function of the number of galaxies occupying a
single dark matter halo and, to a lesser extent, dynamical
friction. Remarkably, these processes conspire to produce a
correlation function in a flat, $\Omega_0=0.3$, CDM model that is
close to a power-law over nearly four orders of magnitude in
amplitude. This model agrees well with the correlation function of
galaxies measured in the APM survey. On small scales, the model
galaxies are less strongly clustered than the dark matter, whereas on
large scales they trace the occupied halos. Our clustering
predictions are robust to changes in the parameters of the galaxy
formation model, provided only those models that match the bright end
of the galaxy luminosity function are considered.
\end{abstract}


\section{Introduction}
\label{sec:intro}

Studies of the clustering of cosmological dark matter have progressed
enormously in the past twenty years. The dynamical evolution of the dark
matter is driven by gravity and fully specified initial conditions are
provided in current cosmological models. This problem can therefore be
attacked quite cleanly using N-body simulations (see Jenkins et al. 1998,
Gross et al. 1998 and references therein.)  Studies of the clustering
properties of galaxies, on the other hand, are much more complicated
because galaxy formation includes messy astrophysical processes such as gas
cooling, star formation and feedback from supernovae. These processes
couple with the gravitational evolution of the dark matter to produce the
clustering pattern of galaxies. Because of this complexity, progress in
understanding galaxy clustering has been slow. Yet, theoretical modelling
of galaxy clustering is essential if we are to make the most of the new
generation of galaxy redshift surveys, the two-degree field (2dF, Colless
1996) and Sloan Digital Sky Survey (SDSS, Gunn \& Weinberg 1995), and of
the new data on galaxy clustering at high redshift that has been
accumulating recently (e.g. Adelberger et al. 1998, Governato et al. 1998,
Baugh et al. 1999).

Two kinds of simulation techniques are being used to approach galaxy
clustering from a theoretical standpoint. The first of these attempts
to follow galaxy formation by simulating directly dark matter and gas
physics in cosmological volumes (eg. Katz et al. 1992, Evrard et al
1994, Weinberg et al. 1998, Blanton et al 1999, Pearce et
al. 1999). Because the resolution of such simulations is limited,
phenomenological models are required to decide when and where stars
and galaxies form and to include the associated feedback effects. The
advantage of this approach is that the dynamics of cooling gas are
calculated correctly without the need for simplifying assumptions. The
disadvantage is that even with the best codes and fastest computers
available, the attainable resolution is still some orders of magnitude
below what is required to resolve the formation and internal structure
of individual galaxies in cosmological volumes. For example, the gas
resolution element in the large Eulerian simulations of Blanton et
al. (1999) is around half a megaparsec.  Lagrangian hydrodynamic
methods offer better resolution, but even in this case, this is poorer
than the galactic scales on which much of the relevant astrophysical
processes occur.

A different and complementary approach to studying galaxy clustering
is to use semi-analytic models of galaxy formation. In this case,
resolution is generally not a major issue. The disadvantage of this
technique, compared to hydrodynamic simulations, is that, in
calculating the dynamics of cooling gas, a number of simplifying
assumptions, such as spherical symmetry or a particular flow
structure, need to be made (some of these assumptions are tested
against smoothed particle hydrodynamics simulations by Benson et
al. 1999). As in the direct simulation approach, a model for star
formation and feedback is required. In addition to adequate
resolution, semi-analytic modelling offers a number of advantages for
studying galaxy clustering. Firstly, it is a much more flexible
approach than full hydrodynamic simulation and so the effects of
varying assumptions or parameter choices can be readily
investigated. Secondly, with detailed semi-analytic modelling it is
possible to calculate a wide range of galaxy properties such as
luminosities in any particular waveband, sizes, bulge-to-disk ratios,
masses, circular velocities, etc. This makes it possible to construct
mock catalogues of galaxies that mimic the selection criteria of real
surveys and to investigate clustering properties as a function of
magnitude, colour, morphological type or any other property determined
by the model.

Semi-analytic modelling has been used in two different modalities to study
galaxy clustering. In the first, an analytic model for the clustering of
dark matter halos developed by Mo \& White (1996) is assumed and the
semi-analytic machinery is used to populate halos, generated using
Monte-Carlo techniques, with galaxies. This technique has been extensively
applied by Baugh et al. (1998, 1999). In the second, more direct, approach,
the semi-analytic modelling is applied to dark matter halos grown in a
cosmological N-body simulation. The advantages of this latter strategy are
that it allows a proper treatment of the small scale regime where the Mo \&
White model breaks down and it bypasses any inaccuracies in the analytic
(Press-Schechter) model used to compute the mass function of dark halos in
the pure semi-analytic approach. This technique has been implemented in two
ways. In the simplest case (Kauffmann, Nusser and Steinmetz,
1997, Roukema et al. 1997, Governato et al. 1998), a statistical merger tree
for each halo identified in the N-body simulation is generated in a
Monte-Carlo manner. In the second implementation (Kauffmann et al. 1999a,
199b, Diaferio et al. 1999), the halo merger trees are extracted directly
from the N-body simulation.

In this paper, we adopt the first approach to the combined use of
semi-analytic and N-body techniques (i.e. with Monte-Carlo merger trees)  
to study galaxy clustering. We focus on the specific question
of how the process of galaxy formation couples with the large scale
dynamics of the dark matter to establish the clustering properties of the
galaxy population. We investigate in detail processes that bias galaxies to
form preferentially in certain regions of space. Previous cosmological dark
matter simulations have established that the dark matter in popular CDM
models tends to be more strongly clustered on small scales than the
observed galaxy population (Jenkins et al. 1998, Gross et al. 1998). We
investigate whether the required antibias arises naturally in these
cosmologies. More generally, we compare the predictions of these models
with observations over a range of scales. The techniques that we use are
described in \S\ref{sec:desc}. The clustering properties of galaxies in our
model are presented in \S\ref{sec:2pt}. The various processes that play a
role in determining how the galaxy distribution is biased relative to the
mass are discussed in \S\ref{sec:nature}. In \S\ref{sec:norms}, we show
that our results are robust to changes in model parameters and finally in
\S\ref{sec:discuss} we discuss our main conclusions.

\section{Description of the Model}
\label{sec:desc}

The two techniques that we employ in this paper, N-body simulations and
semi-analytic modelling, are both well established and powerful theoretical
tools. We do not intend to describe them in detail here, but instead refer
the reader to the appropriate sources.

\subsection{Semi-analytic models}

We use the semi-analytic galaxy formation model of Cole et al. (1999) to
populate dark matter halos with galaxies. The merger history of dark matter
halos is followed using a Monte-Carlo approach based on the extended
Press-Schechter formalism (Press \& Schechter 1974; Bond et al.  1991;
Bower 1991; Lacey \& Cole 1993). Within each halo, galaxy formation is
followed using a set of simple, physically-motivated rules that model the
processes of gas cooling, star formation, feedback from supernovae and
stellar evolution. The result is a fully specified model of galaxy
formation with a relatively small number of free parameters which can be
fixed by constraining the model to match the observed properties of the
local galaxy population (e.g. the luminosity function or the Tully-Fisher
relation). Once constrained in this way, the model makes predictions for a
whole range of galaxy properties (e.g. colours, sizes, bulge-to-disk
ratios, rotation speeds etc.) at both the present day and at high
redshift. In this work we extend this list of predictions to include the
spatial clustering of galaxies, in particular the two-point correlation
function.

Whilst our models are similar in principle to those of Kauffmann et al.
(1999a) they use different rules for star formation and feedback and also
include the effects of chemical enrichment due to star formation. We also
choose to constrain the model parameters in a rather different way, as
discussed below.

\begin{figure*}
\begin{tabular}{cc}
\psfig{file=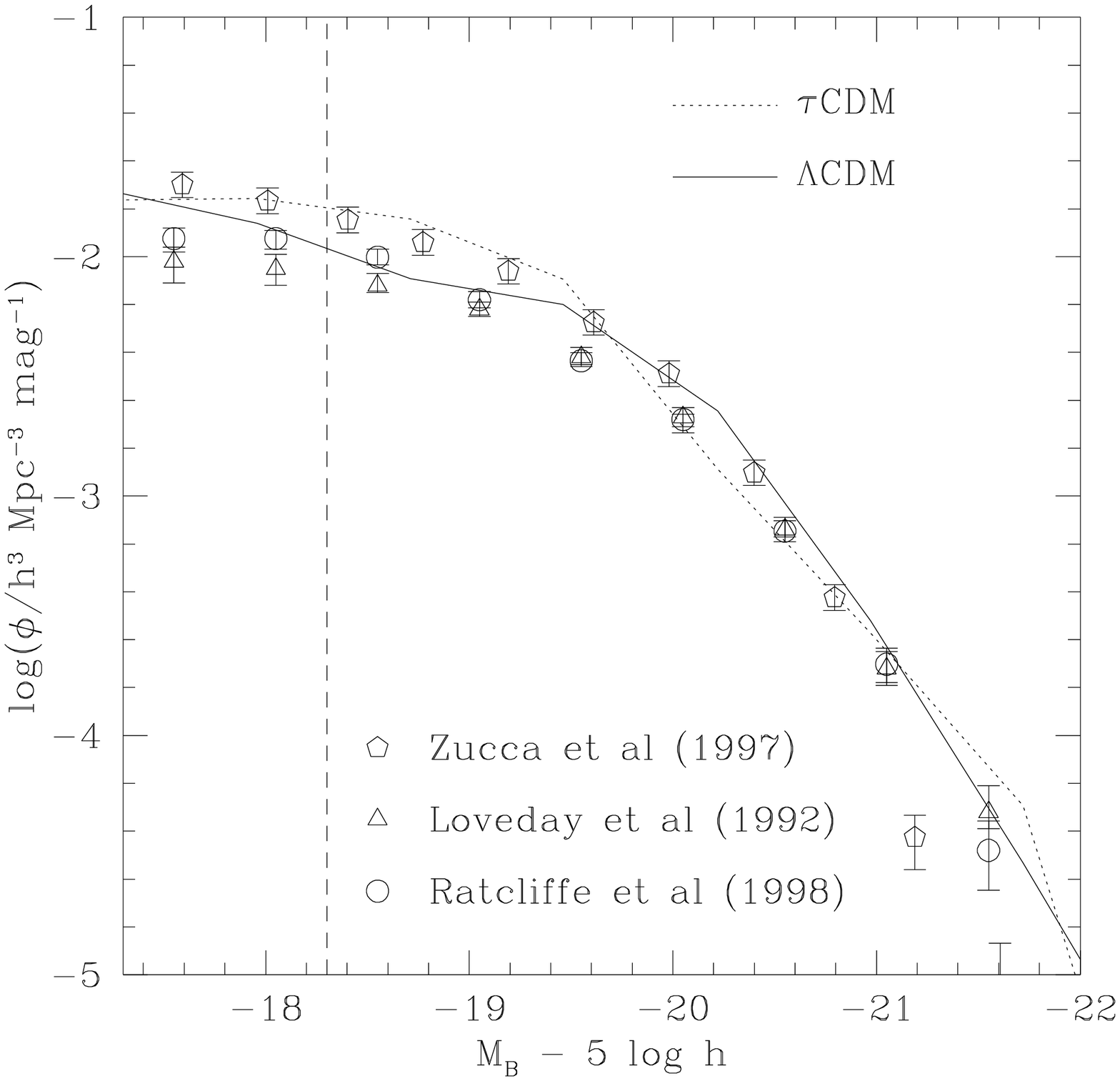,width=80mm} &
\psfig{file=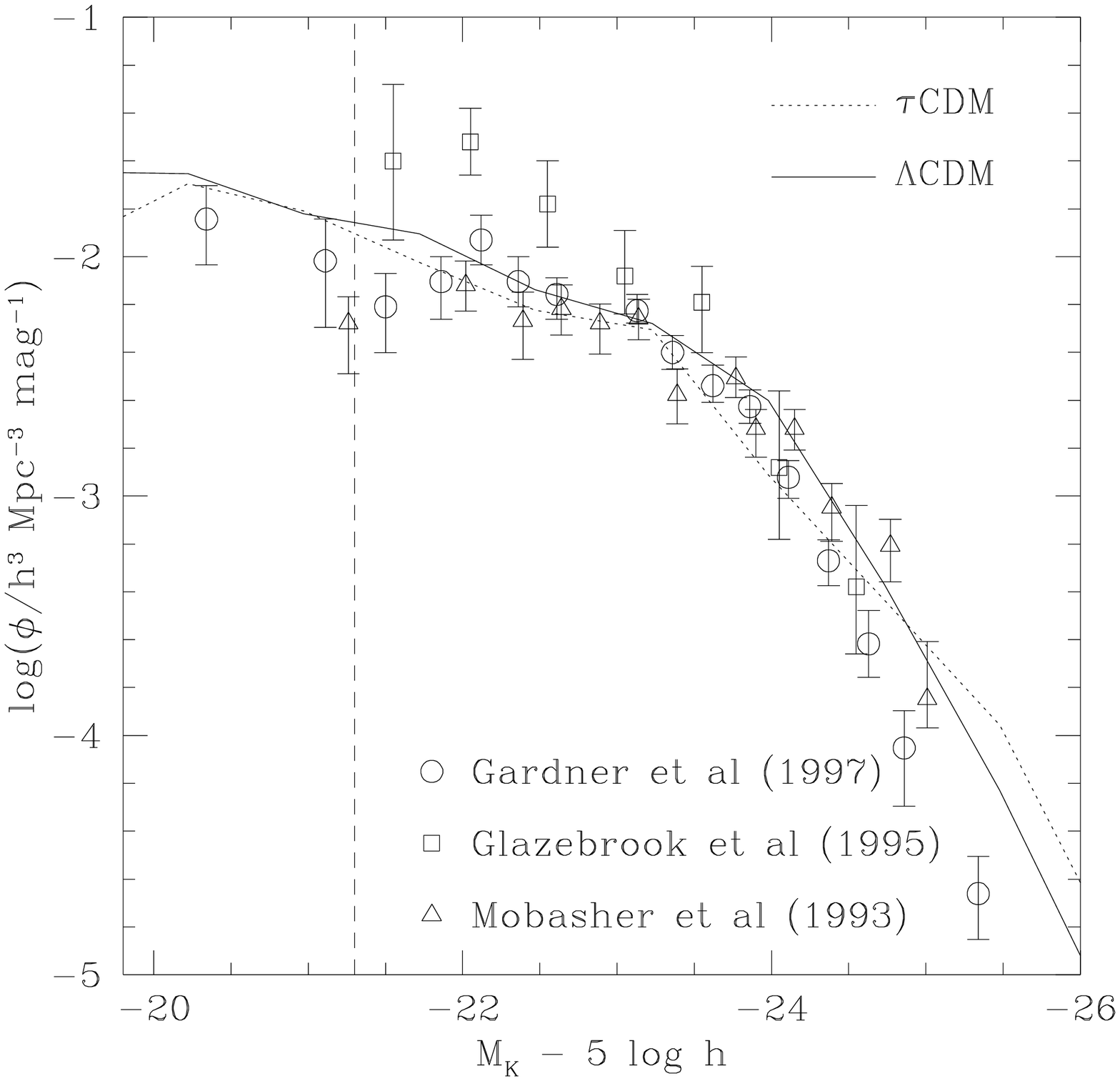,width=80mm}
\end{tabular}
\caption{B and K band luminosity functions for the $\tau$CDM (dotted line)
and $\Lambda$CDM (solid line) reference models. Points with error bars show
a selection of observational determinations of the luminosity
functions. The luminosity functions are shown only as faint as the
$\tau$CDM resolution limit and the vertical dashed line shows the
resolution limit for $\Lambda$CDM models.}
\label{fig:refLF}
\end{figure*}

\subsection{Incorporation into N-body simulations}

We first locate dark matter halos within the simulation volume by use
of a group finding algorithm. This provides a list of approximately
virialised objects within the simulation. For each such halo we
determine the position and velocity of the centre of mass and also
record the positions and velocities of a random sample of particles
within the halo. The list of halo masses from the simulation is fed
into the semi-analytic model of galaxy formation in order to produce a
population of galaxies associated with each halo. The merger tree for
each dark matter halo is generated using a Monte-Carlo method, as
opposed to being extracted directly from the N-body simulation (as was
done by Kauffmann et al. 1999a).

Each galaxy is assigned a position and velocity within its halo. Since the
semi-analytic model distinguishes between central and satellite galaxies,
we locate the central galaxy at the centre of mass of the halo and assign
it the velocity of the centre of mass. Any satellite galaxies are located
on one of the randomly selected halo particles and are assigned the
velocity of that particle. In this way, by construction, satellite galaxies
always trace the density and velocity profile of the dark matter halo in
which they reside. 

Once galaxies have been generated and assigned positions and velocities
within the simulation it is a simple process to produce catalogues of
galaxies with any desired selection criteria (e.g. magnitude limit, colour, 
etc.) complete with spatial information (or, equally simply, with redshift
space positions to enable the study of redshift space distortions).

\subsection{Reference models}

We have made use of the ``GIF'' simulations carried out by the Virgo
Consortium. These are high resolution simulations of cosmological
volumes of dark matter carried out in four different cosmologies:
$\tau$CDM and $\Lambda$CDM (which are used as our reference models),
SCDM and OCDM (which we consider briefly in \S\ref{sec:norms}). These
models are described in detail by Jenkins et al. (1998) and the
simulations are described by Kauffmann et al. (1999a). Briefly, the
simulations model boxes of order 100 $h^{-1}$ Mpc in size with nearly
17 million particles, each of mass approximately $10^{10} h^{-1}
M_{\odot}$. The critical density models (SCDM and $\tau$CDM) have
$h=0.5$ and spectral shape parameter (as defined by Efstathiou, Bond
\& White 1992) $\Gamma = 0.5$ and 0.21 respectively, whilst the low
density models ($\Lambda$CDM and OCDM) have $h=0.7$, $\Omega _0 = 0.3$
and $\Gamma = 0.21$. The $\Lambda$CDM model is made to have a flat
geometry by inclusion of a cosmological constant. All the models are
normalised to produce the observed abundance of rich clusters
today. Dark matter halos were identified using the
``Friends-of-Friends'' algorithm (Davis et al. 1985) with a linking
length of $b = 0.2$; only halos containing 10 or more particles are
considered. The ability to resolve halos of this mass allows us to
determine the properties of galaxies up to one magnitude fainter
than $L_*$.

\begin{figure}
\psfig{file=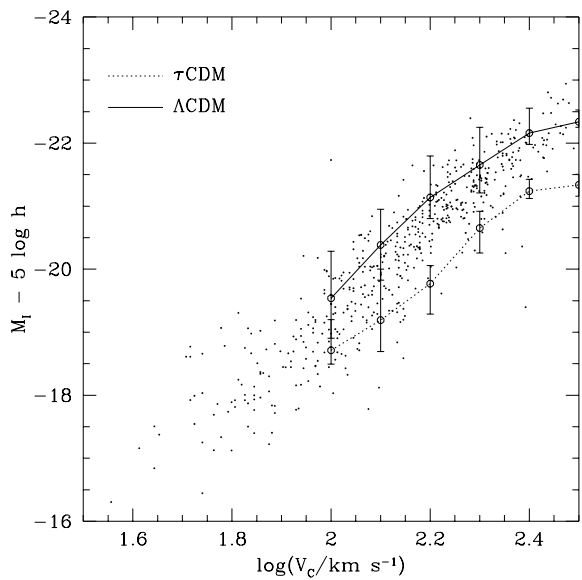,width=80mm}
\caption{Tully-Fisher relations in the $\tau$CDM (dotted line) and
$\Lambda$CDM (solid line) reference models. These models are
constrained by the luminosity function. Points are the observational
data of Mathewson, Ford \& Buchhorn (1992). Each line is plotted using
the circular velocity of each galaxy's dark matter halo and indicates
the median of the distribution, while the error bars indicate the 10\%
and 90\% intervals. Galaxies are selected by their bulge-to-total
ratio in dust-extincted I-band light, which is required to be in the
range 0.02 to 0.24, and are required to have at least 10\% of the mass
of their disk in the form of cold gas.}
\label{fig:refTF}
\end{figure}

We construct two reference semi-analytic models with the same cosmological
parameters as the corresponding GIF simulations. The $\tau$CDM and
$\Lambda$CDM models both reproduce the local B and K-band luminosity
functions, including the exponential cut-off at bright magnitudes,
reasonably well as shown in Fig.~\ref{fig:refLF}. The $\Lambda$CDM model
also produces a close fit to the I-band Tully-Fisher relation constructed
using the circular velocities of the dark matter halos in which they
formed, as may be seen in Fig.~\ref{fig:refTF}. (When the circular
velocities of the galaxies themselves are used instead, the model
velocities are about 30\% too large; see Cole et al. (1999) for a full
discussion.)  In contrast the $\tau$CDM model misses the Tully-Fisher
zero-point by nearly 1 magnitude. (The model Tully-Fisher relations plotted
in Fig.~\ref{fig:refTF} are for galaxies selected by their bulge-to-total
ratio in dust-extincted I-band light, which must lie between 0.02 and
0.24. This approximately matches the range of galaxy types included in the
sample of Mathewson, Ford \& Buchhorn (1992). Furthermore, only galaxies
which have more than 10\% of their disk mass in the form of cold gas are
included. Without a significant fraction of cold gas a galaxy would not
have identifiable spiral structure and measurable HI rotation signal.

The $\Lambda$CDM model is similar to the reference model of Cole et
al. (1999). In both cases, the model parameters were chosen so as to obtain
a reasonable match to a subset of local data, most notably the galaxy
luminosity function. The model used in this paper was selected before the
reference model of Cole et al. (1999) had been fully specified and so there
are small differences in the values of some of the parameters in the two
models. These differences are immaterial for our present purposes. For
example, in a forthcoming paper (Benson et al. 1999, in preparation) we
use the reference model of Cole et al. (1999) to explore further
clustering properties of galaxies. There we show that the two-point
correlation function for galaxies in the reference model differs from the
one presented in this paper only by an amount comparable to the scatter
seen in Fig. \ref{fig:xiLFnorm} (for models which are good fits to the
luminosity function.)

%The small differences compensate for the slightly different values of
%$\sigma_8$ and $\Gamma$ that Cole et al. (1999) use (these parameters
%are essentially fixed for us by the N-body simulation), and for the
%different dark matter halo mass function that we use (i.e. the mass
%function found in the N-body simulation rather than the
%Press-Schechter mass function that Cole et al. employ).

All semi-analytic models considered in this paper include the effects 
of dust on galaxy luminosities calculated using the models of Ferrara et
al. (1999), unless otherwise noted. The model parameters that are
varied in this work are listed in Table \ref{tb:refmods}. The role of
each, and the way in which these parameters are constrained by a set
of observations of the local Universe, are discussed in detail by Cole
et al. (1999). We briefly describe each parameter below:

\noindent \parbox{15.0mm}{$\Omega _b$}\begin{minipage}[t]{70.0mm}Fraction
of the critical density in the form of baryons.\end{minipage}
\parbox{15.0mm}{$\alpha _{\mathrm{hot}}$,
$v_{\mathrm{hot}}$}\begin{minipage}[t]{70.0mm}These determine the strength
of supernovae feedback. Specifically they determine $\beta$, the mass of
gas reheated per unit mass of stars formed, through the relation $\beta =
(v_{\mathrm{hot}}/v_{\mathrm{circ}})^ {-\alpha_{\mathrm{hot}}}$, where
$v_{\mathrm{circ}}$ is the galactic disk circular velocity.\end{minipage}
\parbox{15.0mm}{$\alpha _*$, $\epsilon _*$}\begin{minipage}[t]{70.0mm}These
determine the star formation timescale, $\tau _* = \epsilon _*^{-1} \tau
_{\mathrm{dyn,disk}} (v_{\mathrm{circ}}/200 \hbox{km s}^{-1})^{\alpha _*}$,
where $\tau _{\mathrm{dyn,disk}}$ is the disk dynamical
timescale.\end{minipage}
\parbox{15.0mm}{$f_{\mathrm df}$}\begin{minipage}[t]{70.0mm}This determines
the dynamical friction timescale used to calculate galaxy merger rates
within dark matter halos. The dynamical friction timescale is set equal to
the expression of Lacey \& Cole (1993) multiplied by this
factor.\end{minipage}
\parbox{15.0mm}{$r_{\mathrm core}$}\begin{minipage}[t]{70.0mm}Hot gas in
dark matter halos is assumed to have a density profile given by a
$\beta$-model with $\beta=2/3$. The parameter $r_{\mathrm core}$ is the
core radius expressed in units of the scale length in the dark matter
density profile of Navarro, Frenk \& White (1996). \end{minipage}
\parbox{15.0mm}{$\Upsilon$}\begin{minipage}[t]{70.0mm} The ratio of the
total mass in stars to that in luminous stars. This factor therefore
determines the fraction of stars which are non-luminous (i.e. brown
dwarfs).\end{minipage}
\parbox{15.0mm}{$p$}\begin{minipage}[t]{70.0mm}The yield of
metals.\end{minipage}
\parbox{15.0mm}{$R$}\begin{minipage}[t]{70.0mm}The fraction of mass
recycled by dying stars.\end{minipage}
\parbox{15.0mm}{IMF}\begin{minipage}[t]{70.0mm}The stellar initial mass
function.\end{minipage}\\

As noted in Table \ref{tb:refmods} an artificially low value of $f_{\mathrm
df}$ is required in our $\tau$CDM model in order to obtain a good fit to
the local B and K band luminosity functions. The rapid galaxy merger rate
that results from this choice will deplete the number of galaxies living in
high mass halos, and so may affect the correlation function of
galaxies. However, in \S\ref{sec:LFnorm} we show that altering this
parameter produces no significant change in the model correlation
function.

\begin{table}
\caption{The parameters of our two reference models, using the notation
described in the text.}
\label{tb:refmods}
\begin{tabular}{lcc}
\hline \textbf{Parameter} & \textbf{$\tau$CDM model} & \textbf{$\Lambda$CDM
model} \\ \hline
$\Omega _{\mathrm{b}}$ & 0.08 & 0.02 \\ $\alpha _{\mathrm{hot}}$ & 2.0 &
2.0 \\ $v_{\mathrm{hot}}$ (km/s) & 300.0 & 150.0 \\ $\epsilon _*$ & 0.02 &
0.01 \\ $\alpha _*$ & -0.5 & -0.5 \\ $f_{\mathrm df}$ & 0.1$^{\dag}$ & 1.0
\\
$r_{\mathrm core}$ & 0.1 & 0.1 \\ $\Upsilon$ & 1.23 & 1.63 \\ $p$ & 0.04 &
0.02 \\ $R$ & 0.28 & 0.41 \\ IMF & Salpeter (1955) & Kennicutt (1983) \\
\hline
\end{tabular} \\
$^{\dag}$ As described in Cole et al. (1999) $f_{\mathrm df}$ should
be approximately 1 or larger. Here we use an artificially low value in
order to obtain a good fit to the local B and K band luminosity
functions for the $\tau$CDM model.
\end{table}

\begin{figure*}
\begin{tabular}{cc}
\psfig{file=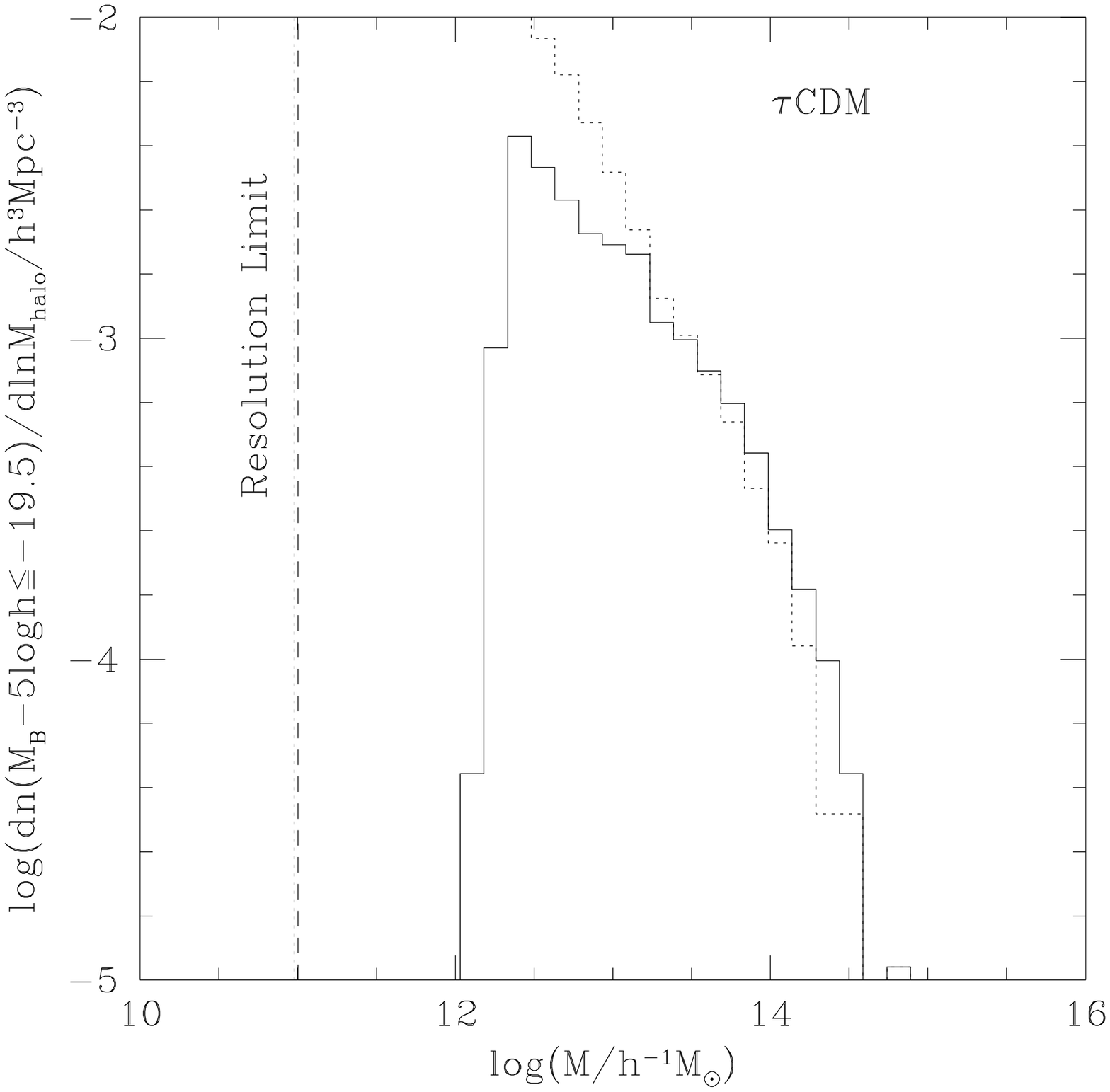,width=80mm} &
\psfig{file=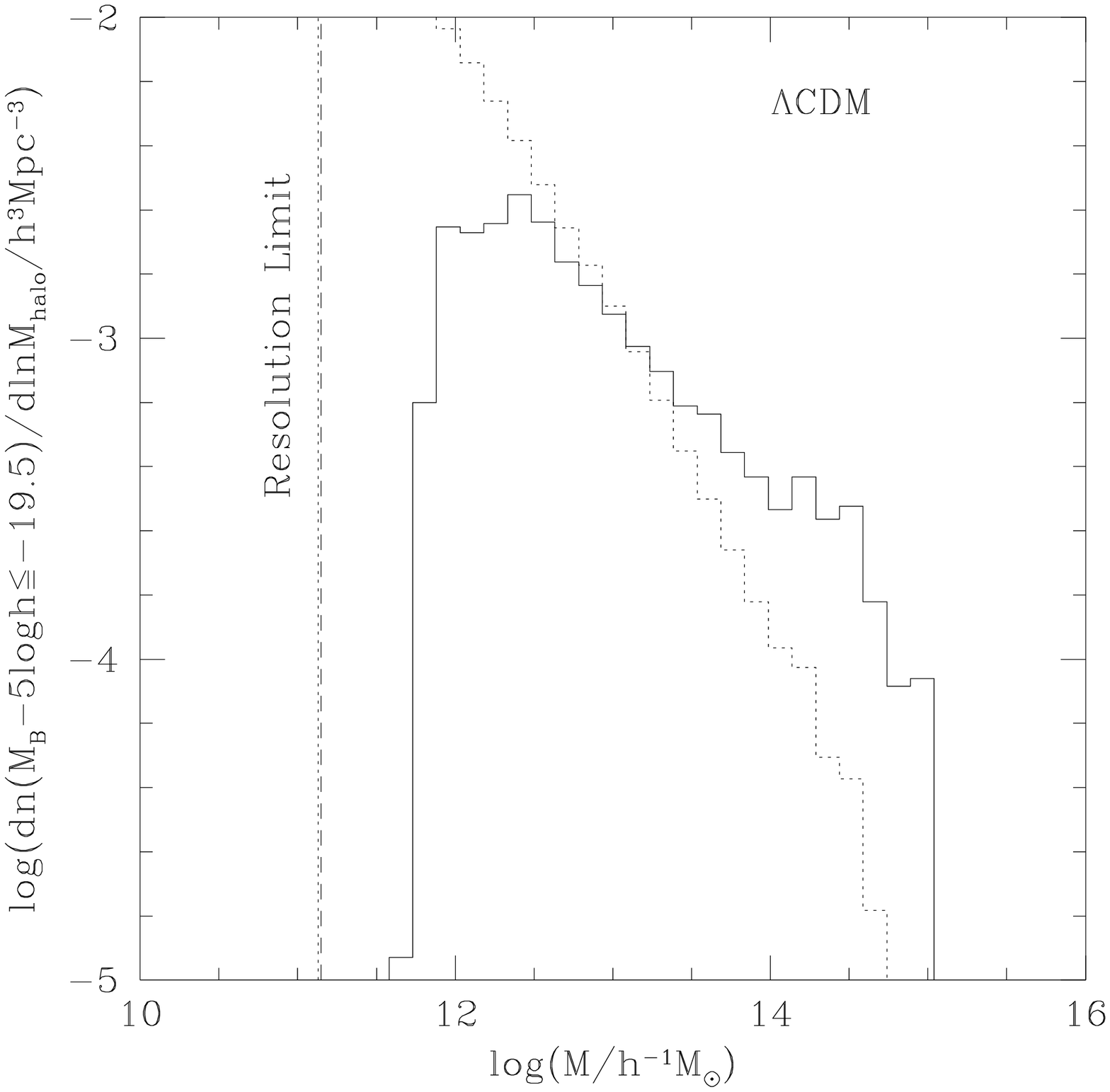,width=80mm}
\end{tabular}
\caption{Mass function of halos containing galaxies with
$M_{\mathrm{B}} - 5 \log h \leq -19.5$ in our $\tau$CDM (left hand
panel) and $\Lambda$CDM (right hand panel) models. Mass functions
weighted by number of galaxies are shown by the solid lines while
unweighted mass functions are shown by dotted lines. These halos are
well above our resolution limit (equal to the mass of a group of 10
particles in each simulation).}
\label{fig:MFref}
\end{figure*}

The smallest halo that can be resolved in the N-body simulation
determines the faintest galaxies for which our model catalogues are
complete. We consider only galaxies brighter than $M_{\mathrm{B}} - 5
\log h = -19.5$ and we have checked, in each case, that the model is
complete to this magnitude. A model is complete if the lowest mass
halo which can contain a galaxy of interest is above the group
resolution limit in the simulation (which is 10 times the particle
mass). Fig.~\ref{fig:MFref} displays the halo mass functions for
galaxies brighter than $M_{\mathrm{B}} - 5 \log h = -19.5$ in our two
reference models and shows that these two models are complete to this
magnitude limit. The minimum mass halo occupied by galaxies is $5.3
\times 10^{11} h^{-1} M_{\odot}$ in $\Lambda$CDM and $1.5 \times
10^{12} h^{-1} M_{\odot}$ in $\tau$CDM. The faintest galaxies which
are fully resolved in the semi-analytic models have $M_{\mathrm{B}} -
5 \log h \approx -18.3$, $M_{\mathrm{K}} - 5 \log h \approx -21.3$ and
$M_{\mathrm{B}} - 5 \log h \approx -17.3$, $M_{\mathrm{K}} - 5 \log h
\approx -19.8$ in $\Lambda$CDM and $\tau$CDM respectively. When
varying model parameters we have checked that the galaxy samples are
complete.

\section{Clustering of galaxies}
\label{sec:2pt}

\subsection{The galaxy two-point correlation function}

The evolution of dark matter in the linear regime is well understood
analytically, and can be followed into the non-linear regime using
N-body simulations (e.g. Jenkins et al. 1998), or theoretically
inspired model fits to the simulation results (Hamilton, Kumar, Lu \&
Matthews 1991; Peacock \& Dodds 1996). The case for galaxies is very
different. Galaxies are generally believed to form near regions of
high density (as in the heuristic ``peaks bias'' model of galaxy
formation; see, for example, Bardeen et al. 1986). If young galaxies
formed only in halos with masses greater than the characteristic
clustering mass, $M_*$ (the mass for which the r.m.s. density
fluctuation in the Universe equals the critical overdensity for
collapse in the spherical top-hat model), at birth they would be
biased with respect to the dark matter. However, galaxy formation is
an ongoing process occurring in a range of halo masses, so any initial
bias will evolve with time.

Several authors (Davis et al. 1985, Tegmark \& Peebles 1998, Bagla 1998)
have shown that if galaxies could be assigned permanent tags at birth, then
their correlation function would approach that of the dark matter at late
times because the clustering due to gravitational instability eventually
becomes much greater than that due to the initial formation sites of
galaxies. They show that this is true even in simple, continuous models of
galaxy formation.

However, the Universe is more complex than this. It is difficult, if not
impossible, to assign a permanent tag to a galaxy since galaxies evolve and
sometimes merge. Therefore, as we look to higher redshifts, it is unlikely
that we will be observing the same population of galaxies that we see at
$z=0$. For example, in a survey with a fixed apparent magnitude limit we
should expect to see the galaxy correlation function initially decreasing
to higher $z$, as the characteristic clustering mass decreases.
Eventually, however, the correlation function should begin to rise as the
apparent magnitude limit selects only the brightest and most massive
galaxies at high redshift which are intrinsically more clustered than the
average galaxy. These points have been discussed in detail by Kauffmann et
al. (1999a) and Baugh et al. (1999). Thus, the apparent evolution of the
galaxy clustering pattern depends on the internal evolution of the galaxies
themselves as well as on the variation of their positions with time. In our
semi-analytic model both of these forms of evolution are explicitly
included.

\begin{figure*}
\begin{center}
\begin{tabular}{cc}
\psfig{file=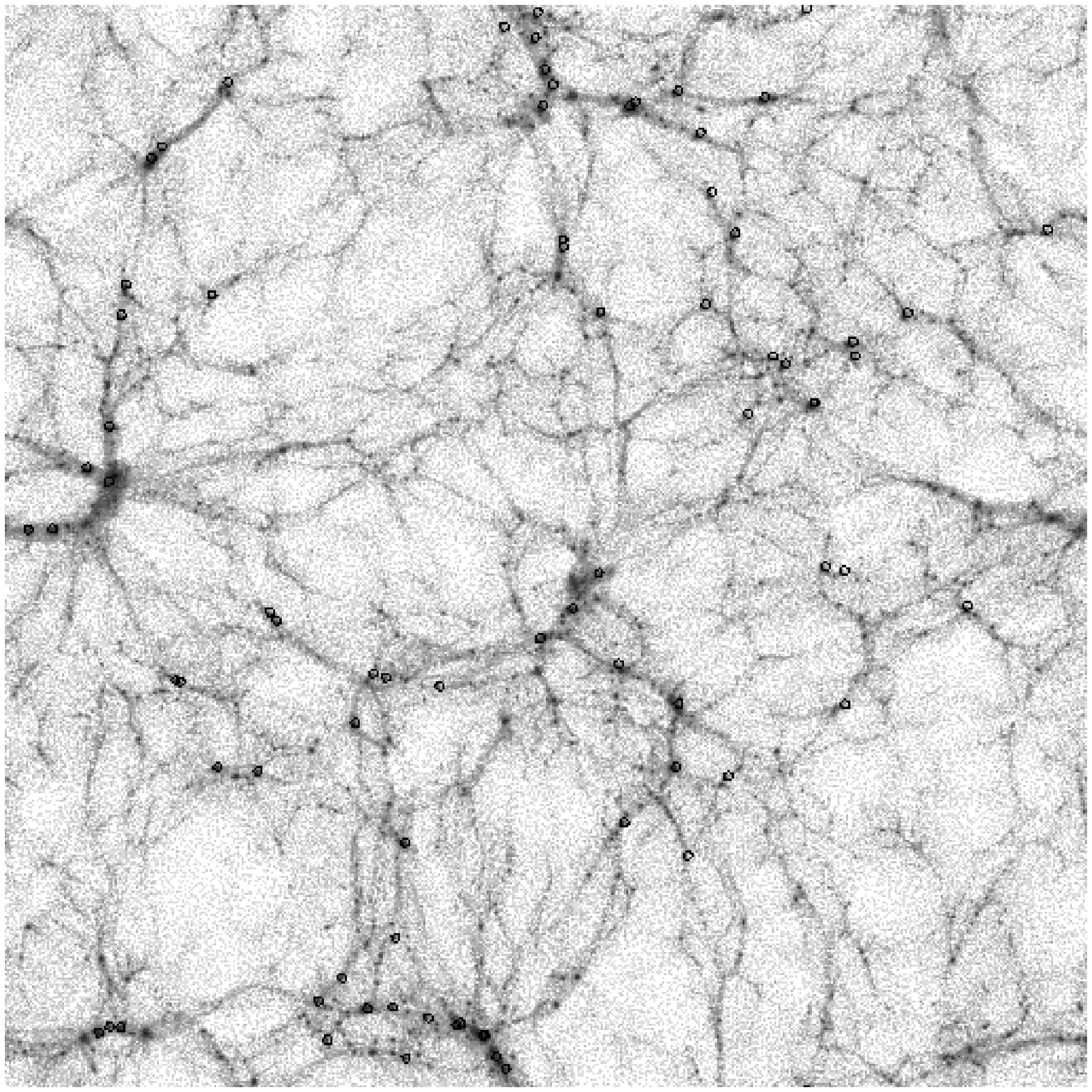,width=80mm} & 
\psfig{file=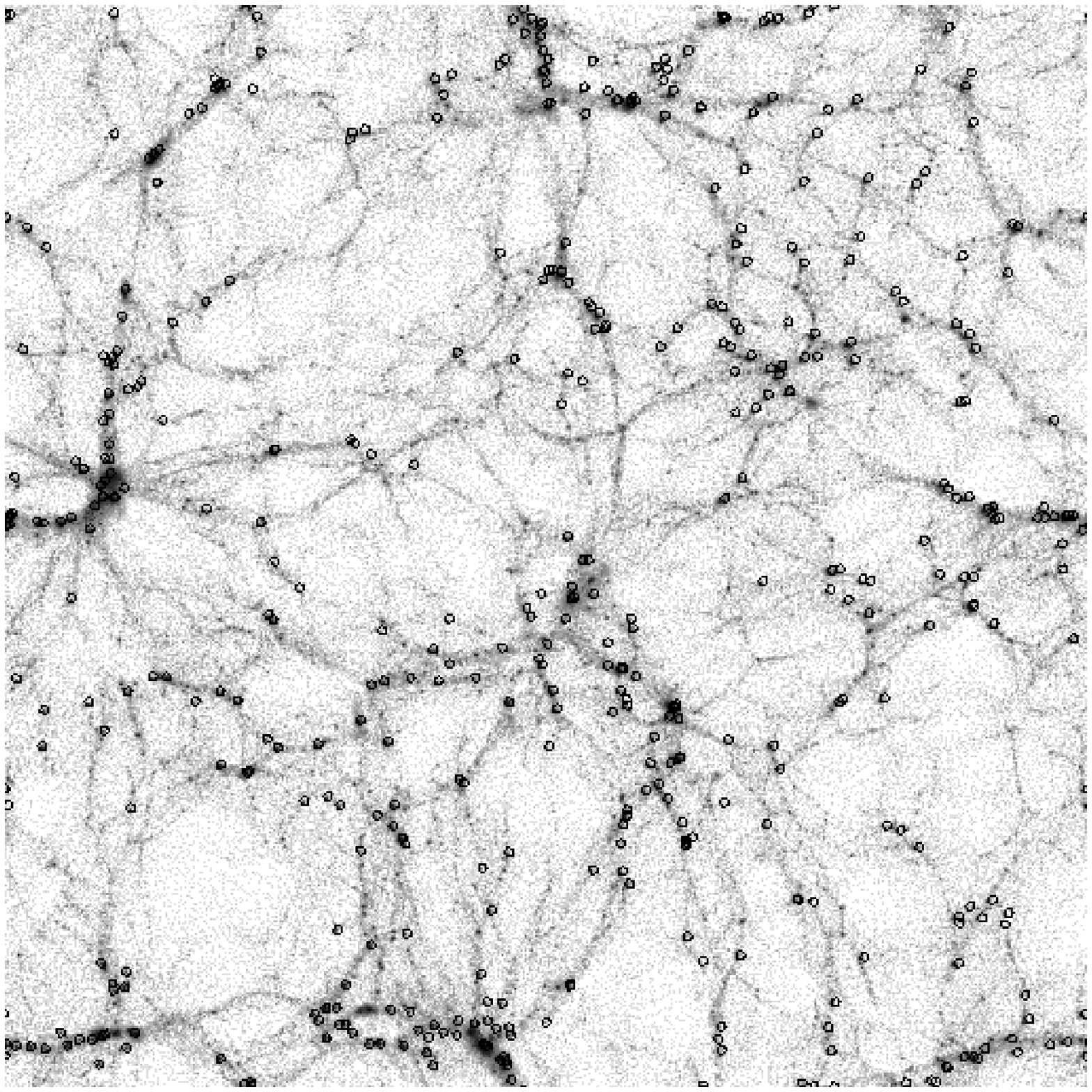,width=80mm}
\\ $\tau$CDM & $\Lambda$CDM
\end{tabular}
\caption{The left hand panel shows the locations of galaxies brighter than
$M_{\mathrm{B}} - 5 \log h = -19.5$ in a $\tau$CDM model. The figure shows
a slice of the dark matter simulation $85 \times 85 \times 4.7 h^{-3}$
Mpc$^3$ in size. The density of dark matter is indicated by the greyscale
(with the densest regions being the darkest). Overlaid are the positions of
the galaxies, indicated by open circles. The right hand panel shows the
equivalent slice from a $\Lambda$CDM model (the GIF simulations all have
the same phases, hence the similarity of the structure), the slice in this
case being $141 \times 141 \times 8 h^{-3}$ Mpc$^3$ in size.}
\label{fig:greys}
\end{center}
\end{figure*}

\begin{figure*}
\begin{center}
\begin{tabular}{cc}
\psfig{file=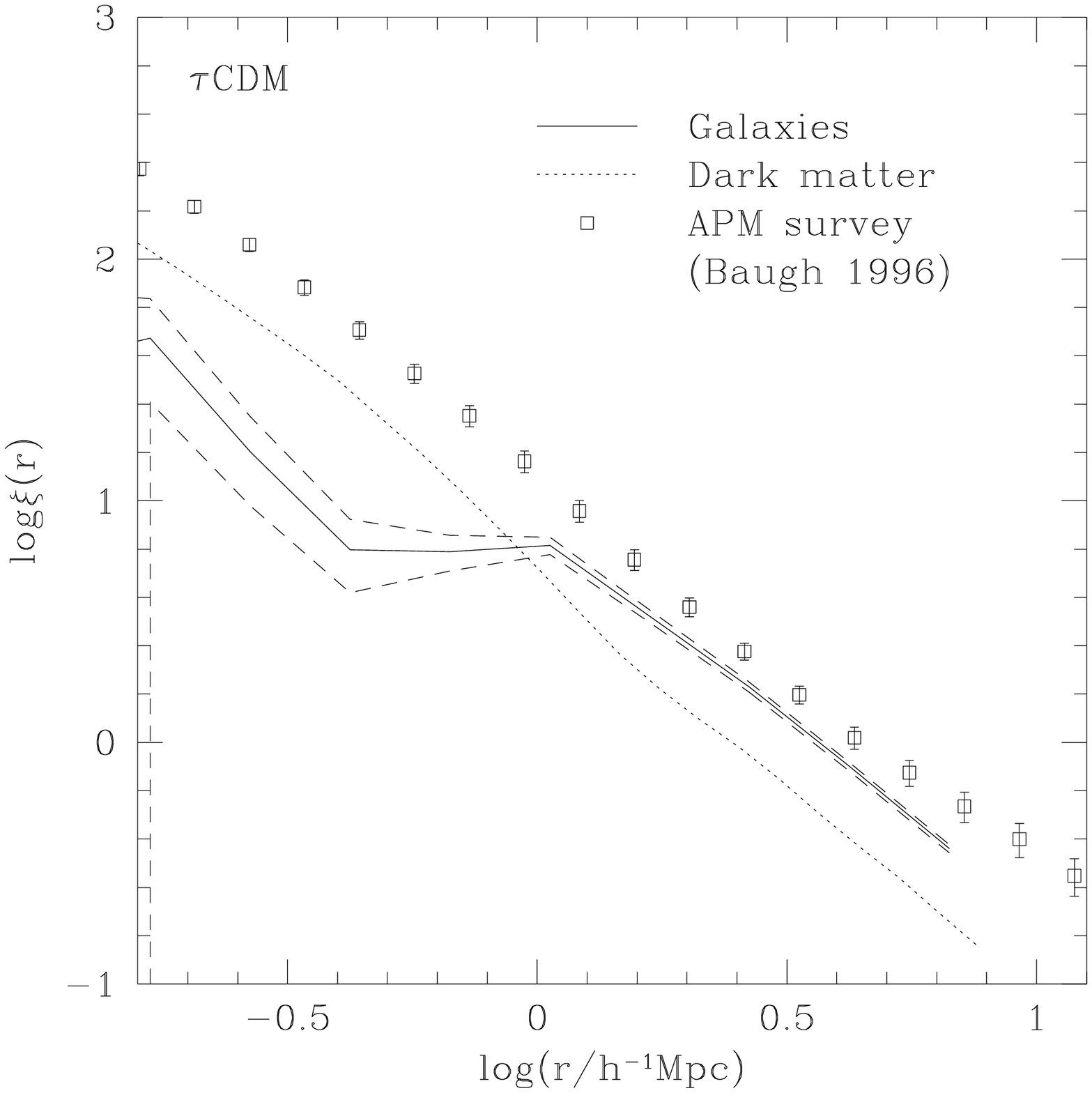,width=80mm} &
\psfig{file=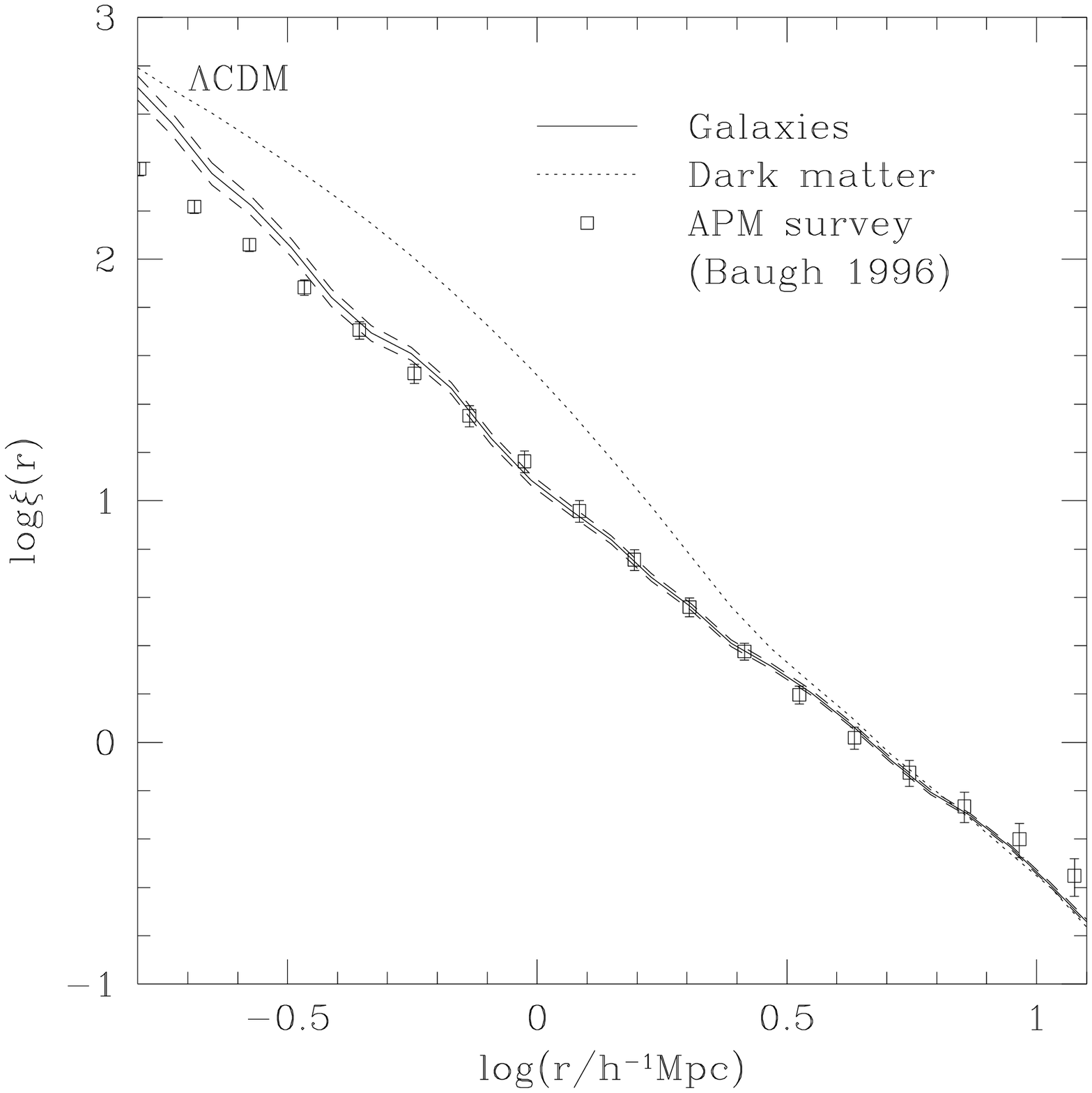,width=80mm}
\end{tabular}
\caption{The left hand panel shows the two-point correlation function of
galaxies brighter than $M_{\mathrm{B}} - 5 \log h = -19.5$ in a $\tau$CDM
model as a solid line. The dashed lines to either side indicate the Poisson
sampling errors. This is compared to the observed APM real-space
correlation function (points with error bars) and to the mass correlation
functions in the N-body simulations (dotted line). The right hand panel
shows the equivalent plot for a $\Lambda$CDM model.}
\label{fig:xiref}
\end{center}
\end{figure*}

The results of the techniques described in the previous section are shown
in Fig.~\ref{fig:greys} \& \ref{fig:xiref}. Fig.~\ref{fig:greys} shows
slices through the GIF $\Lambda$CDM and $\tau$CDM dark matter simulations
on which we have overlaid the positions of galaxies from our models. The
galaxies can be seen to trace out structure in the dark matter and to avoid
the underdense regions in the dark matter distribution. The galaxies
clearly follow the large scale structure of the dark matter, but as we will
show, they are biased tracers of the mass. The most obvious difference
between the two diagrams is the smaller number of galaxies in the $\tau$CDM
model. This is simply due to the smaller volume of the $\tau$CDM slice
(approximately five times smaller than the $\Lambda$CDM slice), since the
number of galaxies per unit volume is constrained to be very similar in
each model by the requirement that they match the observed luminosity
function.

Fig.~\ref{fig:xiref} shows the two-point correlation functions of the
model galaxies, and compares them to the observed APM correlation
function (in real-space) and to the correlation function of the
underlying dark matter. The two models show distinct differences in
their behaviour. Most obviously, the $\Lambda$CDM model is very close
to the APM data from $r \approx 0.3 h^{-1}$ Mpc to $r \approx 10
h^{-1}$ Mpc, whilst the $\tau$CDM model fails to achieve a large
enough amplitude on scales $\geq 1.0 h^{-1}$ Mpc and drops even
further below the observed correlation function on smaller scales. The
$\tau$CDM model shows a strong bias on large scales. The bias
parameter, defined as the square root of the ratio of the galaxy and
mass correlation functions, is approximately 1.4. The $\Lambda$CDM
model, on the other hand, is essentially unbiased on large scales. Both
models show an anti-bias on smaller scales. It is interesting to note
that the galaxy correlation functions do not display the same features
as the dark matter. For example the shoulder in the $\Lambda$CDM dark
matter correlation function at 3 $h^{-1}$ Mpc is not present in the
galaxy correlation function. Instead, the latter is remarkably close
to a power-law form over about four orders of magnitude in amplitude.

\subsection{Systematic effects}

In this section we consider two systematic effects which may affect the
clustering properties of galaxies in our models: dynamical friction in
groups and clusters and our procedure for constructing merger trees for the
dark matter halos. We show that neither of these significantly affects the
two-point correlation function.

\subsubsection{Dynamical friction}
\label{sec:dynfric}

Our models do not accurately account for the effects of dynamical
friction on the spatial position of satellite galaxies in halos. The
simulations lack the resolution to follow this process directly. We
can, however, correctly model the two extremes of this effect. If the
dynamical friction timescale is much longer than the age of the halo,
then the galaxy orbit is close to its original orbit, and so our
placement scheme, consisting of identifying galaxies with randomly
chosen halo particles, is correct on average. Conversely, if the
dynamical friction timescale were much shorter than the halo lifetime,
the satellite galaxy would have sunk to the bottom of the halo
potential well and merged with the central galaxy. This effect is
included in the semi-analytic model. Therefore, it is only in the
intermediate range where the dynamical friction timescale is of the
same order as the halo lifetime that our models do not accurately
reproduce the galaxy positions within clusters.

To estimate the effect of dynamical friction on the correlation function we
have tried perturbing the galaxy positions using the following simple
model. From the calculation of the dynamical friction timescale in an
isothermal halo given by Lacey \& Cole (1993) (their equation B4), it can
be seen that the orbital radius of a galaxy in a circular orbit, $r$,
decays with time as

\begin{equation}
r = r_{\mathrm{i}} \sqrt{ {t_{\mathrm{df}} - t \over t_{\mathrm{df}}}},
\end{equation}

\noindent where $r_{\mathrm{i}}$ is the initial orbital radius of the
galaxy when the halo forms, at $t = 0$, and $t_{\mathrm{df}}$ is the
dynamical friction timescale of the galaxy, given by Lacey \& Cole
(1993).  Here, to mimic this behaviour, each satellite galaxy is first
assigned a position in the halo tracing the dark matter as before and
then its distance from the halo centre is reduced by a factor
$r/r_{\mathrm{i}}$.

\begin{figure}
\begin{center}
\psfig{file=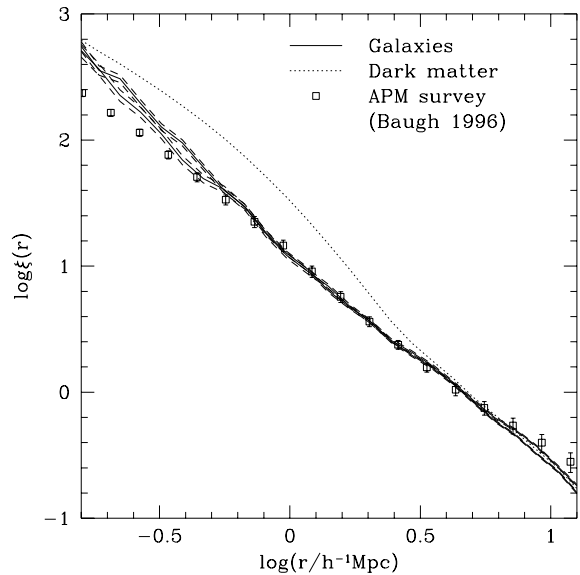,width=80mm}
\caption{The correlation function in our $\Lambda$CDM reference model, with
and without the effects of dynamical friction on satellite galaxy
positions. The thin solid line shows the standard model (the dashed lines
indicating the Poisson errors), whilst the thick lines show the same model
with an estimate of dynamical friction effects included.}
\label{fig:dynfric}
\end{center}
\end{figure}

Fig.~\ref{fig:dynfric} shows the correlation function in our
$\Lambda$CDM reference model with and without this dynamical friction
effect included. Dynamical friction causes only a slight increase in
the clustering amplitude on small scales, $< 0.5 h^{-1}$ Mpc, (since
galaxies are drawn closer together inside halos). However the effect
is small and can be safely neglected.

\subsubsection{Merger tree construction}

\begin{figure}
\psfig{file=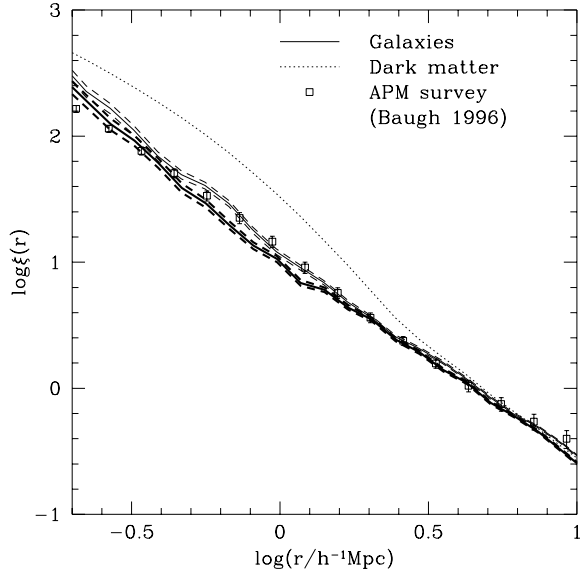,width=80mm}
\caption{A comparison of correlation functions for galaxies brighter than
$M_{\mathrm{B}} - 5 \log h = -19.5$ in our $\Lambda$CDM reference model
with the results of a model with an artificial mass resolution designed to
mimic the models of Kauffmann et al. (1999a). The low resolution model is
shown as a thick solid line whilst our reference model is shown by thin
solid lines. The dashed lines indicate the Poisson sampling errors.}
\label{fig:lowres}
\end{figure}

As noted in \S\ref{sec:intro}, one difference between this work and that of
Kauffmann et al. (1999a) is that they extract merger trees for dark matter
halos directly from the N-body simulation, whereas we extract the final
mass of the halo and generate the merger tree using the extended
Press-Schechter Monte-Carlo formalism. There are advantages to both
techniques. Extracting the halo trees from the simulation circumvents any
possible discrepancy between the extended Press-Schechter predictions and
the merging histories in the N-body simulation, although it has been shown
that the two are statistically equivalent (see for example Lacey \& Cole
1994; Lemson \& Kauffmann 1999; Somerville et al. 1998). In particular,
Lemson \& Kauffmann (1997) have studied the statistical properties of halo
formation histories in N-body simulations and find no detectable dependence
of formation history on environment, as expected in the Press-Schechter
theory. Thus, the fact that we construct merger trees similarly for halos
in high and low-density regions should make little or no difference to our
results.  Furthermore, since here we are only interested in the statistical
properties of the galaxy population, our approach is justified. 

One drawback of the direct extraction technique is that the merging trees
become limited by the resolution of the simulation. Like us, Kauffmann et
al. identified halos containing at least 10 particles. Since this mass
resolution limit applies at all times in the simulation, such a halo cannot
have been formed by merging, as it might have done in a higher resolution
simulation. Furthermore, even large mass halos might have significantly
modified merging histories due to this artificial resolution limit. The
analytic merging trees that we generate do not suffer from this problem.
The effective mass and time resolutions can be made as small as desired,
until convergence is reached. We demonstrate the effects of the resolution
limit by considering a $\Lambda$CDM model in which we artificially impose
an effective mass resolution equivalent to that in the Kauffmann et al.
models. Fig.~\ref{fig:lowres} shows that the differences between the models
with and without the artificial mass resolution limit are, in general
insignificant, although there is a region (approximately from separations
of 0.4 to 1.5 $h^{-1}$ Mpc) where the disagreement between the two is
significant.

Finally, it should be noted that since we extract the final masses of halos
from the N-body simulation, our models do not suffer from the
well-documented (but small)  differences between the Press-Schechter and
N-body mass functions at low mass (e.g. below $\sim 10^{14} h^{-1}
M_{\odot}$), discussed by Efstathiou, Frenk, White \& Davis (1988), Lacey \&
Cole (1994) and Somerville et al. (1998). 

\section{The nature of bias}
\label{sec:nature}

\begin{figure}
\psfig{file=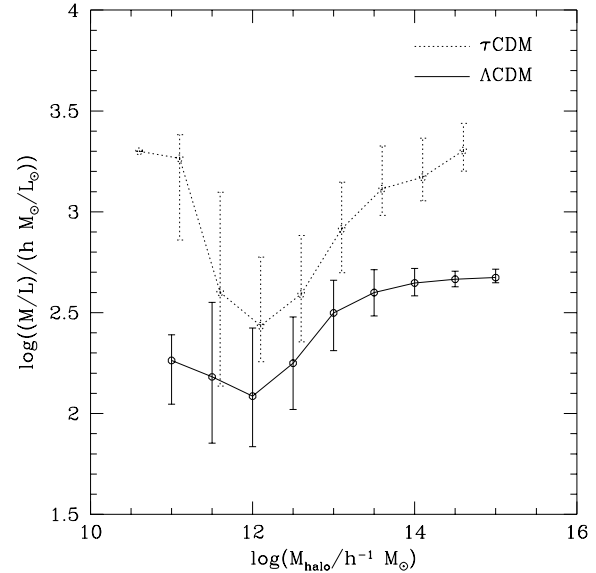,width=80mm}
\caption{The B-band mass-to-light ratio of halos in our models. The
dotted line corresponds to $\tau$CDM and the solid line to
$\Lambda$CDM.  Lines show the median mass-to-light ratio, whilst the
error bars indicate the 10 and 90 percentiles of the distribution. For
reference, the mean mass-to-light ratio in the simulation as a whole
is about 1440 and 470 $h M_{\odot}/L_{\odot}$ in the $\tau$CDM and
$\Lambda$CDM cosmologies respectively, with an uncertainty of about
20\% due to unresolved galaxies.}
\label{fig:MLLF}
\end{figure}

In the models explored here, galaxies do not trace the mass exactly
because galaxy formation proceeds with an efficiency which depends
on halo mass. In the lowest mass halos, feedback from
supernovae prevents efficient galaxy formation, whilst in the high
mass halos, gas is unable to cool efficiently by the present day
thereby inhibiting galaxy formation. These effects can be seen in the
mass-to-light ratios (in the B-band) of halos in our reference models
plotted in Fig.~\ref{fig:MLLF}. The mass-to-light ratio is strongly
dependent on halo mass.  Initially, it decreases as halo mass
increases, before turning upwards and levelling off at close to the
universal value for the highest mass halos in the simulations. The
minimum, at around $10^{12} h^{-1} M_{\odot}$, marks a preferred mass
scale at which the efficiency of galaxy formation is greatest. The
mass-to-light ratio varies by more than a factor of 3 over the range
of masses plotted here. As a result of this varying mass-to-light
ratio we expect a complex, scale-dependent bias to arise and this is,
in fact, seen in our two reference models. The clustering of galaxies
is controlled by the intrinsic bias of their host halos, the
non-linear dynamics of the dark matter and the processes of galaxy
formation.

\begin{figure*}
\begin{tabular}{cc}
\psfig{file=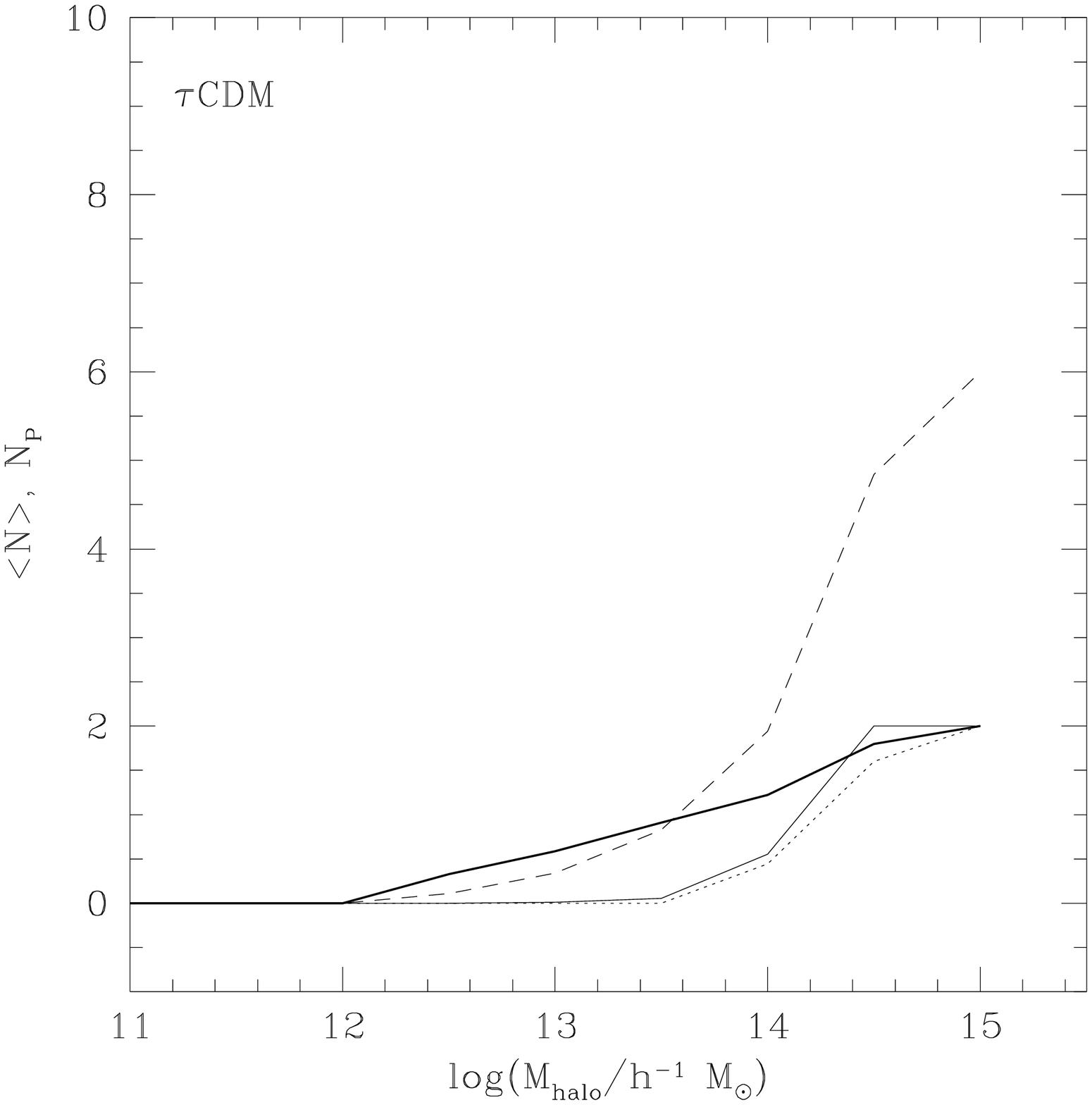,width=80mm} &
\psfig{file=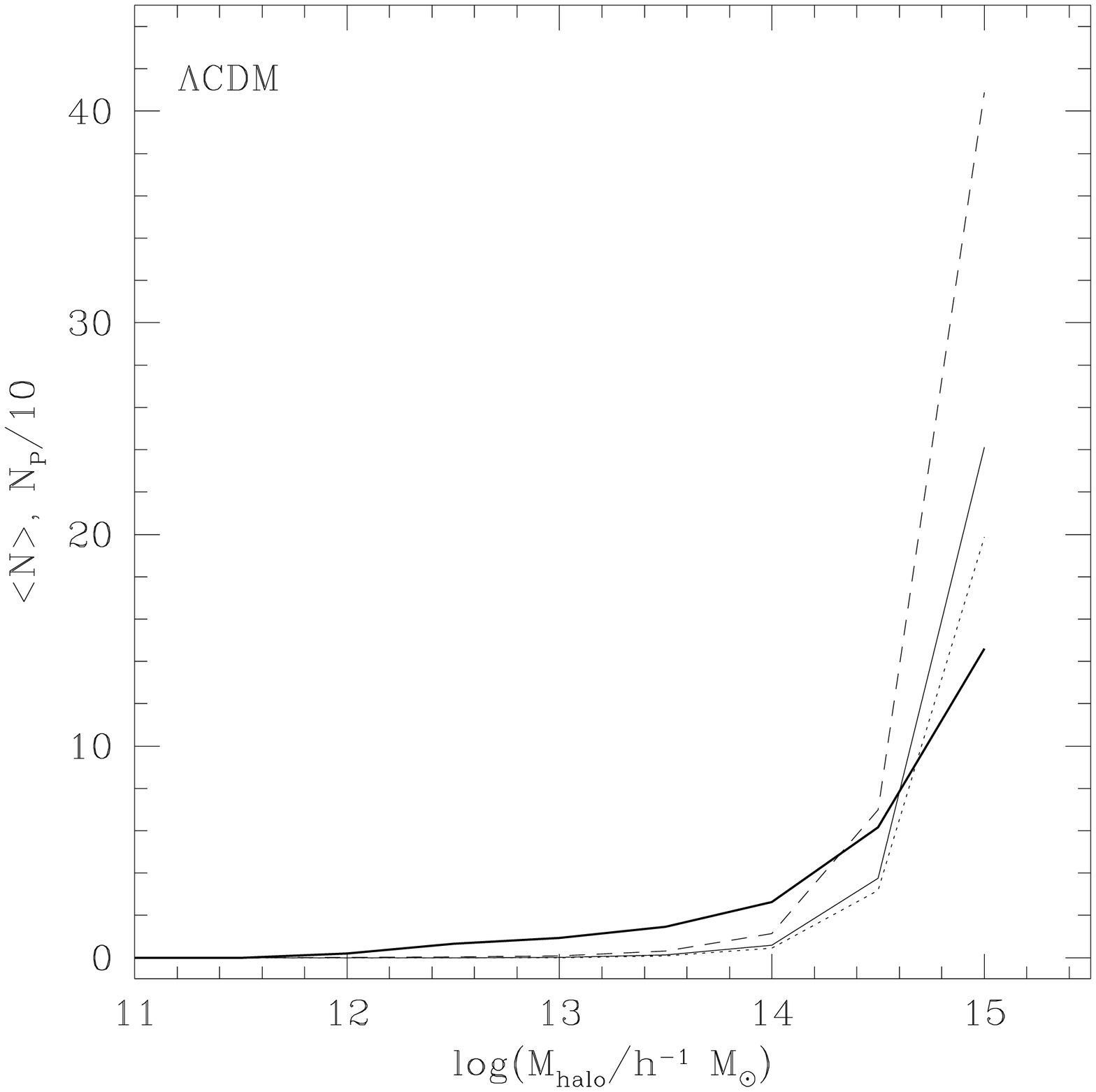,width=80mm}
\end{tabular}
\caption{The mean number of galaxies, brighter than $M_{\mathrm{B}} - 5
\log h = -19.5$ per halo as a function of halo mass. The plots are for the
$\tau$CDM (left hand panel) and the $\Lambda$CDM (right hand panel)
models. Note that unoccupied halos are included when computing the mean.
The thick solid line shows the mean number, $\bar{N}$, of galaxies per
halo. The remaining lines indicate the mean number of galaxy pairs per halo
as defined by Equation \ref{eq:pairs}, for three different probability
distributions, $P(N;M)$: ``true'' (thin solid line); ``average'' (dotted
line) and ``Poisson'' (dashed line). Note the different scales in the two
plots.}
\label{fig:NMref}
\end{figure*}

Fig.~\ref{fig:NMref} shows the mean number of galaxies per halo as a
function of halo mass in our two models. (For future reference we also
plot the mean number of pairs of galaxies per halo as defined by
Equation \ref{eq:pairs} below.) Below the $10^{12.5} h^{-1} M_{\odot}$
and $10^{13} h^{-1} M_{\odot}$ bins in the $\tau$CDM and $\Lambda$CDM
models respectively, halos always contain zero or one galaxy (i.e. the
number of pairs is zero). This is simply because there is not enough
cold gas in the halo to form two or more galaxies of the required
luminosity by the present day. At higher halo masses there is a trend
of increasing number of galaxies per halo. The average occupation
increases less rapidly than the halo mass, indicating once again that
the halo mass-to-light ratio increases with increasing mass.

Galaxies brighter than some given absolute magnitude only form in halos
above a certain mass, $M_{\mathrm h}$. On scales much larger than the radii
of these halos, the correlation function of these galaxies will be
proportional to that of the dark matter, with some constant, asymptotic,
large-scale bias, as has been shown by Mo \& White (1996). Behaviour of
this type is seen in both of our reference models. This large scale bias
can be estimated by averaging the Mo \& White analytic bias for all halos
of mass greater than $M_{\mathrm h}$, weighting by the abundance of those
halos and by the number of galaxies residing (on average) within them (see
Baugh et al. 1999).

On smaller scales the situation is more complex. The Mo \& White
calculations break down on scales comparable to the pre-collapse
(Lagrangian) radius of the host halos. If halos of mass $M$ have a
Lagrangian radius $R$, then we expect a reduction of the correlation
of these halos on scales $\leq R$, since these objects must have
formed from spatially exclusive regions of the universe. Halos may
have moved somewhat after their formation and so will not be
completely exclusive below this scale. However, they must be
completely exclusive below their post-collapse (virial) radius
($R_{\mathbf{vir}}$), since no two halos can occupy the same region of
space. Galaxies, however, resolve the internal structure of the halos
and so we should not necessarily expect the same degree of anti-bias
on sub-$R_{\mathrm vir}$ scales in the galaxy distribution although
these exclusion effects may still be apparent to some extent. Instead,
the correlation function will begin to reflect the distribution of
dark matter within the halos since, in our models, galaxies always
trace the halo dark matter.

However, this is still not the whole picture. If $N(M)$ is the average
number of galaxies per halo of mass $M$, then we can define a mass $M'
> M_{\mathrm h}$, where $N(M')=1$ (here $M_{\rm h}$ is the minimum
mass of a halo that can host a galaxy brighter than $M_{\rm B}-5 \log h
= -19.5$). We find $M' = 10^{13}$ and $10^{12} h^{-1} M_{\odot}$ for
the $\tau$CDM and $\Lambda$CDM models respectively.  In halos less
massive than $M'$, we typically find at most a single galaxy and so
the distribution of dark matter within these halos is not resolved.
Instead, our galaxy catalogue contains information only about the
position of the halo centre. In general, the clustering will depend
upon $P(N;M)$, the probability of finding $N$ galaxies in a halo of
mass $M$. In particular, the small-scale clustering will depend upon
the mean number of pairs per halo, which is itself determined by the
form of the $P(N;M)$ distribution. Since the correlation function is a
pair weighted statistic it gives extra weight to distributions with a
tail to high $N$.

\begin{figure*}
\begin{tabular}{cc}
\psfig{file=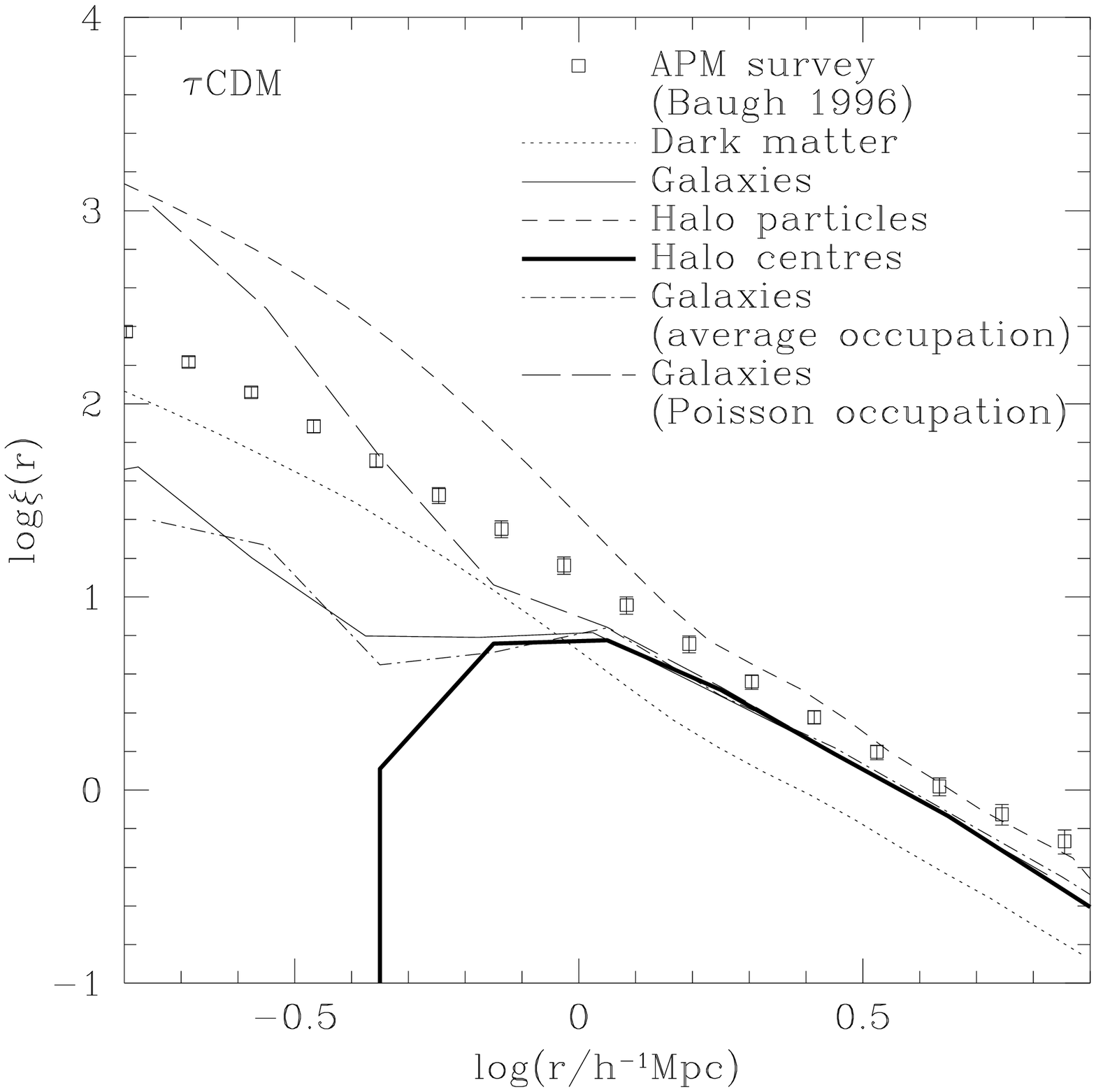,width=80mm} &
\psfig{file=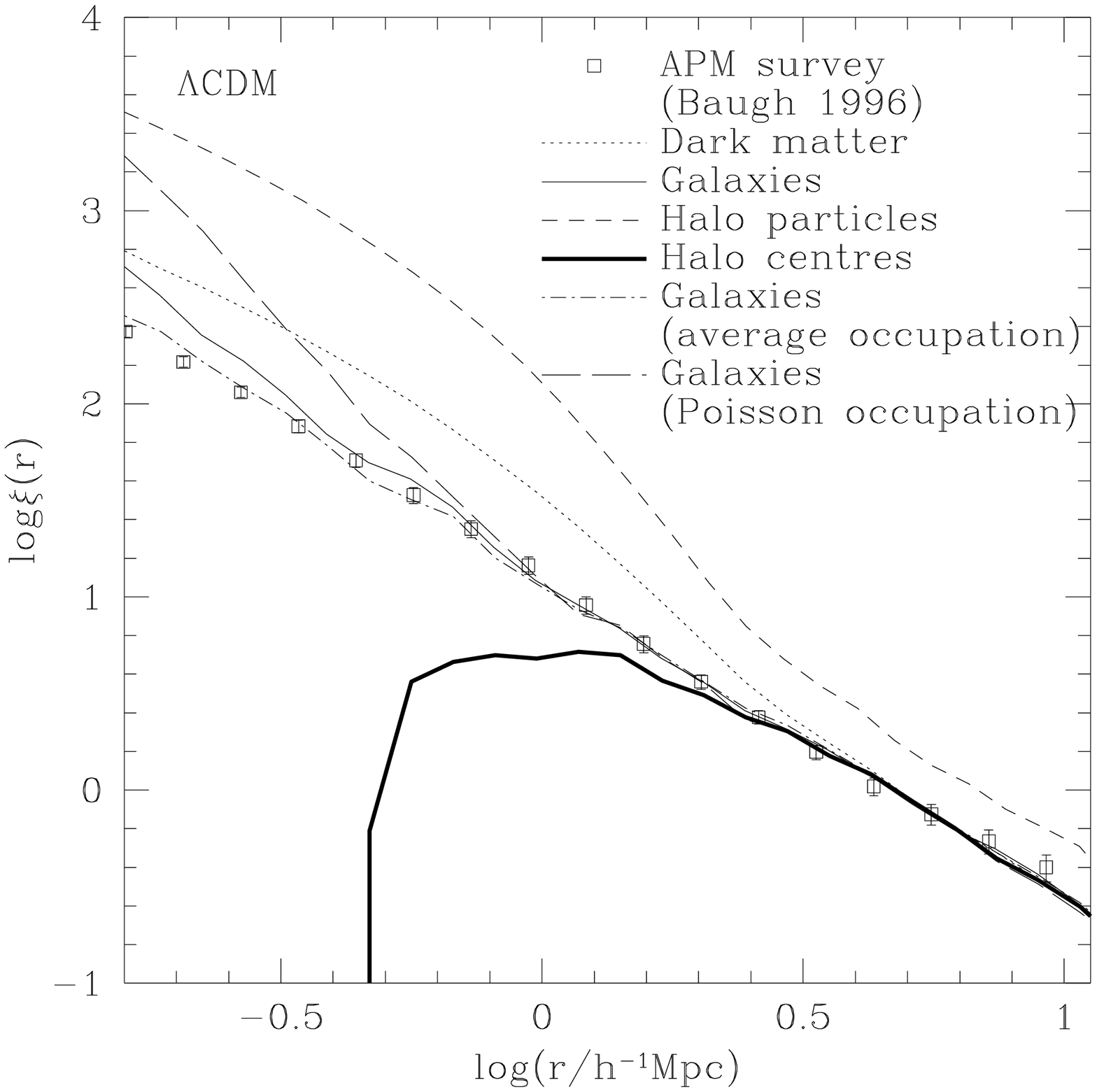,width=80mm}
\end{tabular}
\caption{Correlation functions constructed from different samples of dark
matter particles compared to the observed and model galaxy correlation
functions in the $\tau$CDM reference model (left-hand panel). The various
curves, labelled in the legend, are described in detail in the text. The
right hand panel shows the same plots for the $\Lambda$CDM reference
model.}
\label{fig:reducedDM}
\end{figure*}

The correlation function of galaxies is thus the result of a complex
interplay of several effects: (i) asymptotic constant bias on large
scales; (ii) spatial exclusion of halos; (iii) the number of galaxies
per halo which controls whether the internal structure of the halos is
resolved or not and (iv) the form of the $P(N;M)$ distribution (as
this determines the mean number of pairs of galaxies per halo), which
we discuss below. It is difficult, therefore, to construct an
empirical model that reproduces the results of our full semi-analytic
plus N-body models. It is, however, instructive to plot several
correlation functions which act as bounds on the true galaxy
correlation function.

Fig.~\ref{fig:reducedDM} shows the correlation functions of galaxies in our
model (thin solid line), dark matter in the simulation (dotted line), and
observed galaxies in the APM survey, as measured by Baugh (1996) (squares
with error bars). The short-dashed line is computed from all dark matter
particles that are part of halos of mass greater than $M'$ (i.e.  halos
sufficiently massive to contain galaxies at least some times). This curve
is highly biased with respect to the full dark matter distribution, a fact
that is not surprising given that it excludes the least clustered mass. We
would expect the galaxy correlation function to be similar to this if the
number of galaxies per halo were drawn from a Poisson distribution with
mean proportional to the halo mass, that is, if the mass-to-light ratio
were independent of halo mass. Evidently this is not the case. (The
asymptotic bias of this correlation function is greater than that of the
model galaxies (thin solid line), as weighting by halo mass gives more
weight to the highly biased, most massive halos than does weighting by
galaxy number.) The heavy solid line is the correlation function of halo
centres, with each centre weighted by the model $P(N;M)$ distribution. The
spatial exclusion of halos is evident, causing this curve to drop below
that of the galaxies and finally to plummet to $\xi (r) = -1$ at a scale
comparable to twice the virial radius of the smallest occupied halos.

The dot-dash line shows the correlation function found by placing in
each halo the average number of galaxies per halo of each mass $\left[
\sum _{N=1}^{\infty} N P(N;M) \right]$, using our usual placement
scheme (i.e.  the first galaxy is placed at the halo centre and the
others are attached to random particles in the halo). We refer to this
as the ``average'' model. Obviously, we cannot place the average
number per halo if this is not an integer. In this case, we place a
number of galaxies equal to either the integer immediately below or
immediately above the actual mean with the relative frequencies needed
to give the required mean, which results in a small scatter in the
occupation. Finally, the long-dashed line shows the correlation
function obtained when the number of galaxies in a halo is drawn from
a Poisson distribution with the same mean as the model distribution
(the ``Poisson'' model).

\begin{figure*}
\begin{tabular}{cc}
\psfig{file=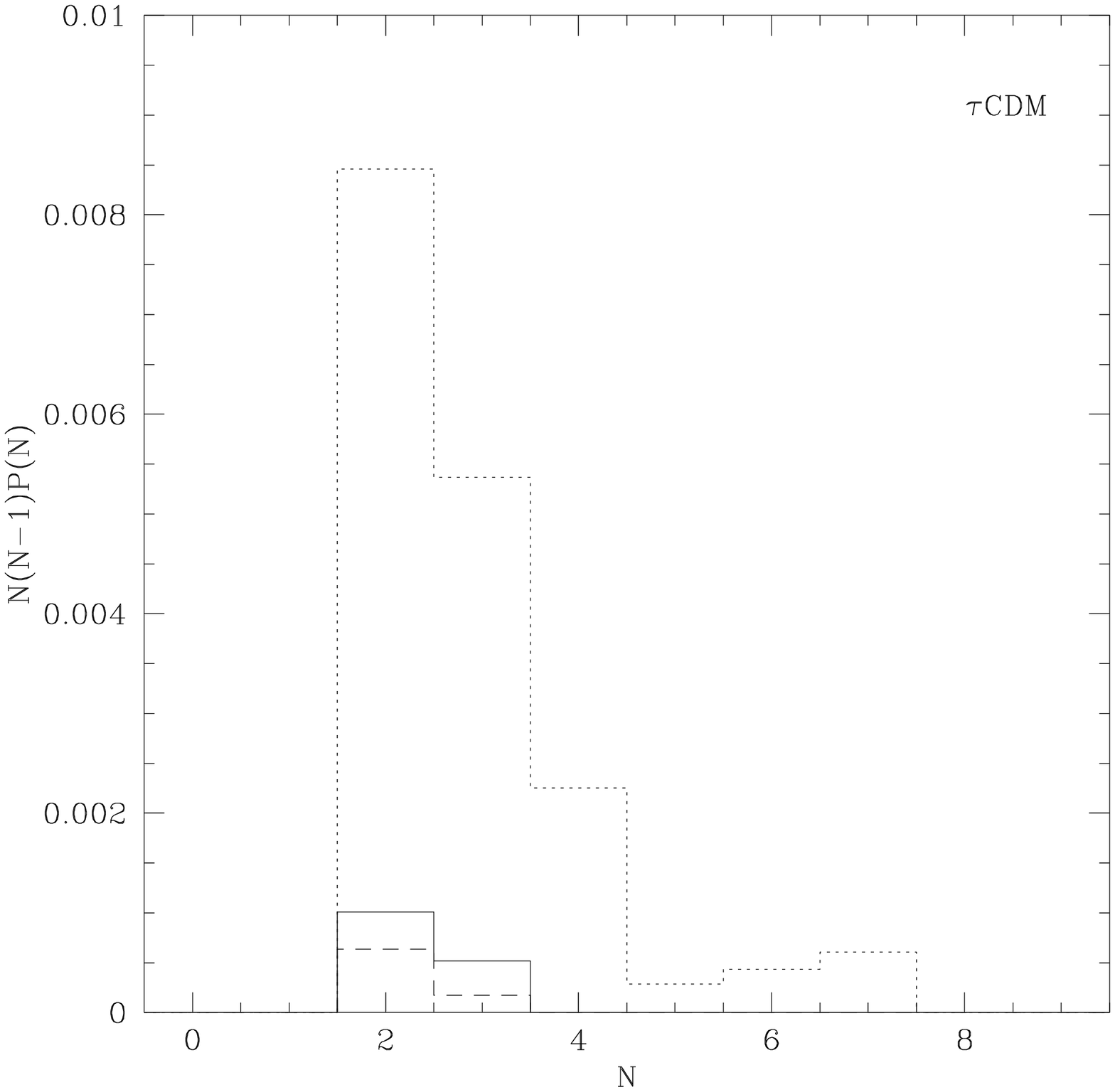,width=80mm} &
\psfig{file=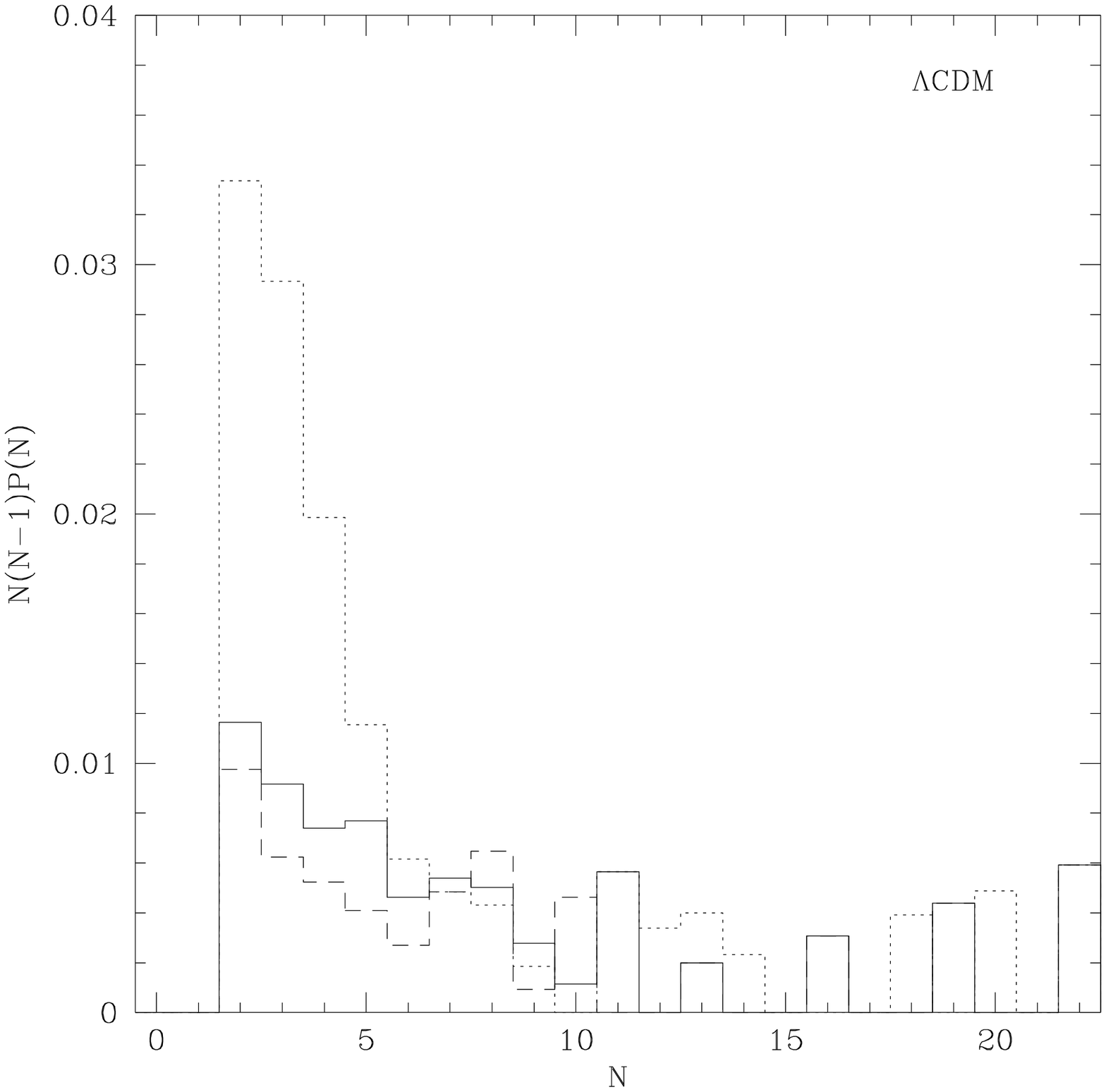,width=80mm}
\end{tabular}
\caption{The probability, $P(N)$, of occupation by $N$ galaxies
(multiplied by $N(N-1)$ for clarity) for halos in the $\tau$CDM model
(left hand panel) and $\Lambda$CDM model (right hand panel). All halos
resolved in the simulations are considered. The solid line shows the
distribution from the actual model, the dotted line shows the
``Poisson'' model distribution and the dashed line shows the
``average'' model distribution. Note that the $N=0$ and $N=1$ bins are
always zero because we choose to weight $P(N)$ by $N(N-1)$.}
\label{fig:occu}
\end{figure*}

The differences between the correlation functions of the full
semi-analytic model and models in which halos are occupied according
to a Poisson distribution or simply with the average galaxy number
(thin solid line, long-dashed line and dot-dash line respectively)
must be due entirely to the form of $P(N;M)$ (i.e. the frequency with
which a halo of a given mass is occupied by $N$ galaxies), since all
these models are, by construction, identical in all other respects,
including the mean halo occupation number. Fig.~\ref{fig:occu}
illustrates the difference between the actual distribution of galaxies
in all halos resolved in the simulations,

\begin{equation}
P(N) = \int ^{\infty} _{M_{\rm min}} P(N;M) n(M) dM \left/ \int ^{\infty} _{M_{\rm min}} n(M)
dM, \right.
\end{equation}

\noindent and the ``Poisson'' and ``average'' models. Here $M_{\rm
min}$ is the mass of the smallest halo that can be resolved in the
simulation. The values of $P(N)$ in this plot are multiplied by
$N(N-1)$ so that the area under the histogram gives the mean number of
pairs per halo. The number of galaxies present in a halo is related to
the structure of the merger tree for that halo. Although the merger
tree is generated by a Monte-Carlo method, this does not produce a
Poisson distribution of progenitor halos. Furthermore, whilst the
Poisson distribution always possesses a tail to arbitrarily high
numbers, the real distribution cannot as there is only enough cold gas
in any one halo to make a limited number of bright galaxies.

Table \ref{tb:pairweights} gives the mean number of pairs found within
a single halo in the two reference models for all halos resolved in
the simulation. This is given by

\begin{equation}
N_{\mathrm{P}} = \sum _{i=0}^{\infty} i(i-1) {\int ^{\infty}_{M_{\rm min}} P(i;M)
n(M) dM \over \int ^{\infty} _{M_{\rm min}} n(M) dM},
\label{eq:pairs}
\end{equation}

\noindent where $P(i;M)$ is the distribution of occupancies
(normalised such that $\sum _{i=0}^{\infty} P(i;M) = 1$) for halos of
mass $M$ and $n(M)$ is the abundance of halos of mass $M$. The number
of pairs in the ``true'', ``average'' and ``Poisson'' distributions is
shown in Figure \ref{fig:NMref}. Note that if we consider two such
halos separated by some distance $\sim r$ then the mean number of
pairs at separation $\sim r$ is

\begin{equation}
\sum _{i=0}^{\infty} \sum _{j=0}^{\infty} ijP(i;M)P(j;M) = \bar{i}^2
\end{equation}

\noindent and $\bar{i}$ is constrained to be equal in all three
distributions.

\begin{table}
\begin{center}
\caption{The mean number of pairs per halo, $N_{\mathrm{P}}$,
calculated for three different distributions of halo occupancy, all
with the same mean. ``Average'' has the same number of galaxies in all
halos in a given mass range (or as close to this distribution as
possible if the mean is not an integer), ``true'' has the distribution
found in our reference models and ``Poisson'' has a Poisson
occupation.}
\label{tb:pairweights}
\begin{tabular}{lccc}
\hline \textbf{Model} & \textbf{average} & \textbf{true} & \textbf{Poisson}
\\ \hline
$\tau$CDM & 0.0005 & 0.0010 & 0.0116 \\
$\Lambda$CDM & 0.0538 & 0.0678 & 0.1262 \\ \hline
\end{tabular}
\end{center}
\end{table}

Thus, we can understand the difference in the clustering amplitudes of
the three correlation functions at small scales (they all agree within
the errors at large scales) simply on the basis of the form of their
$P(N;M)$ function which determines the mean number of pairs per halo,
$N_{\mathrm P}$. For distributions with the same mean, the one with
the lowest number of pairs per halo, $N_{\mathrm P}$, will have the
lowest clustering amplitude, whilst the one with the largest number of
pairs will have the highest clustering amplitude. (In the case of the
``average'' and ``true'' distributions, the correlation functions are
very similar on small scales as the contribution from pairs of
galaxies within a single halo is small compared to that from pairs in
distinct halos.) The consequence of this is that the amplitude of the
small scale end of the observed correlation function tells us
something interesting about $P(N;M)$, namely that it has fewer pairs
than a Poisson distribution and is in reasonable agreement with the
distribution predicted from our semi-analytic model. Thus, the
behaviour of the small separation end of the correlation function is
determined by the physics of galaxy formation. We have checked that
our choice of placing central galaxies at the centre of mass of their
halo does not affect these results. If instead each central galaxy is
placed on a randomly chosen dark matter particle (i.e. if treated just
like a satellite galaxy) the correlation function is unaltered
within the error bars.

The $\tau$CDM reference model shows a break on sub-Mpc scales. As can be
seen in Fig.~\ref{fig:reducedDM} this coincides with the turnover in the
correlation function of halo centres. This turnover is reflected in the
galaxy correlation function because in this model there are too few bright
galaxies in cluster halos to resolve adequately their internal
structure. In the $\Lambda$CDM models, on the other hand, halos are
adequately resolved and the galaxy correlation function remains almost a
power-law, even though that of halo centres turns over. To remove this
feature from the $\tau$CDM model would require more bright galaxies to form
in cluster halos. However, this would have to be accomplished without
significantly increasing the number of bright galaxies in lower mass halos
since these would quickly come to dominate the asymptotic bias which would
therefore become lower than its present value thus exacerbating the
discrepancy between model and observations at large separations. We have
been unable to find a model constrained to match the local luminosity
function which succeeds in removing the sub-Mpc feature in our $\tau$CDM
model whilst simultaneously producing the required asymptotic bias.

\begin{figure}
\psfig{file=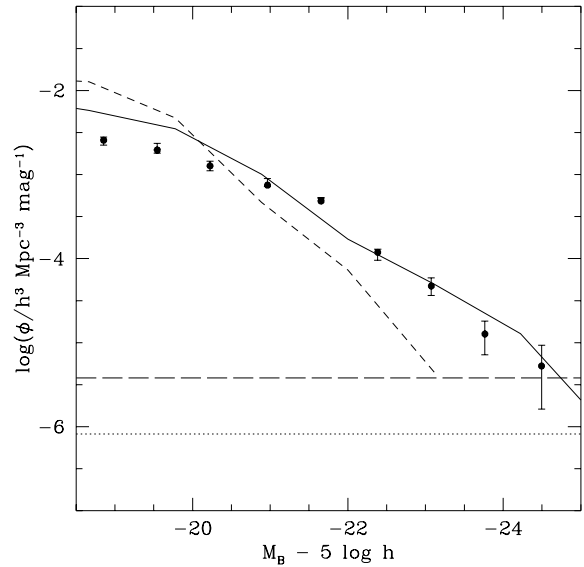,width=80mm}
\caption{The luminosity function of all galactic systems.  Results for
our $\tau$CDM model are shown with the short-dashed line and for our
$\Lambda$CDM model with the solid line. The symbols with error bars
are the observational data from Moore, Frenk \& White (1993). The
horizontal lines indicate the abundance below which the probability of
finding one or more such objects in the entire volume of the
simulation is less than 10\% in the $\tau$CDM (long dashed line) and
$\Lambda$CDM (dotted line) models.}
\label{fig:refgLF}
\end{figure}

The reason for the differences between the two reference models is
illustrated in Fig.~\ref{fig:refgLF}, where we compare our models to
an observational determination of the ``luminosity function of all
galactic systems.''  This function, estimated by Moore, Frenk \& White
(1993), gives the abundance of halos as a function of the {\it total}
amount of light they contain, regardless of how it is shared amongst
individual galaxies. This quantity is difficult to determine
observationally, since one must establish which galaxies are in the
same dark matter halo. Moore, Frenk \& White (1993) approached this
problem by analysing the ``CfA-1'' galaxy redshift survey
\cite{davis82,huchra83} using a modified friends-of-friends group
finding algorithm which was allowed to have different linking lengths
in the radial and tangential directions to account for redshift-space
distortions. These linking lengths were also allowed to vary with
distance, to reflect the changing number density of galaxies in the
survey. It is entirely possible, however, that in some instances this
technique may have grouped together galaxies which actually reside in
distinct dark matter halos. Finally, Moore, Frenk \& White (1993) made
a correction to the luminosity of each identified group to account for
the light from unseen galaxies (i.e. those below the magnitude limit
of the survey). This was done assuming a universal form for the galaxy
luminosity function. Such a form may not, in fact, be applicable to
the real Universe, and is not guaranteed to arise in our models.

Bearing these caveats in mind, we see in Fig.~\ref{fig:refgLF} that
the $\Lambda$CDM reference model and the data are in excellent
agreement except at the faint end where the discrepancy reflects the
fact that the data come from the CfA-1 Survey which has a flatter
luminosity function than the ESO Slice Project (ESP) luminosity function we used to
constrain our semi-analytic model.  By contrast, the $\tau$CDM model
fails to match the luminosity function of all galactic systems, in
spite of the fact that it agrees quite well with the bright end of the
galaxy luminosity function (c.f. Fig.~\ref{fig:refLF}). This model
does not make enough bright galaxies in high mass, highly clustered
halos and it makes too many in low mass, weakly clustered halos. It is
not surprising therefore that the $\tau$CDM galaxy correlation
function falls below the observed data on all scales
(c.f. Fig.~\ref{fig:xiLFnorm}). A similar conclusion applies to the
standard $\Omega _0 =1$ CDM model although in this case the
disagreement with the observed correlation function on large scales is
even worse than in the $\tau$CDM model.  Matching the luminosity
function of all galaxy galactic systems is, therefore, an important
prerequisite for a model to match the two-point correlation function.

\section{Testing the robustness of the predictions}
\label{sec:norms}

The semi-analytic model of galaxy formation is specified by several
parameters. These determine the cosmological model and control
astrophysical processes such as star formation, supernovae feedback and
galaxy merging. Whilst these parameters can be constrained by requiring the
model to reproduce certain local observations (such as the B and K-band
luminosity functions; see Cole et al. 1999), we wish to explore here what
effect altering these parameters has on our estimate of the correlation
function.

Thus, we alter the parameters of the reference model one at a time. We try
to preserve as good a match as possible to the local B-band luminosity
function by giving ourselves the freedom of adjusting the value of
$\Upsilon$ so that the model B-band luminosity function has the correct
amplitude at $L_*$. Since the reference models give a good match not only
to the B-band luminosity function, but also to a variety of other
observational data (such as the distribution of colours, sizes, star
formation rates, etc.), the modified models will, in general, not be as
good as the reference models. Furthermore, in some cases matching the
$M_{\mathrm{B}} - 5 \log h = -19.6$ point of the ESP luminosity function
requires $\Upsilon < 1$ which is unphysical (as it implies negative mass in
brown dwarfs). However, this is not a serious concern here since we are
only interested in testing the robustness of clustering properties to
changes in model parameters. We also consider a few models in which
$\Upsilon$ is set so as to match the zero-point of the I-band Tully-Fisher
relation, rather than the amplitude of the luminosity function at
$L_*$. These are closer to the models of Kauffmann et al. (1999a).

The semi-analytic model with the altered parameters is used to
populate the N-body simulation with galaxies. We then measure the bias
of galaxies brighter than $M_{\rm B}-5 \log h=-19.5$ in each model.

%We then determine the best fit power-law $\xi (r) = (r/r_0)^{-
%\gamma}$ to the correlation function over the range $r = 0.2 h^{-1}$
%Mpc to $r = 10.0 h^{-1}$ Mpc. In the case of the $\tau$CDM cosmology
%the correlation functions are not well fit by a power-law. The power-
%law fit in this case acts simply as a convenient measure of the
%differences in each model.

\subsection{Models constrained by the luminosity function}
\label{sec:LFnorm}

Both of our reference models which are constrained to match the
$M_{\mathrm{B}} - 5 \log h = -19.6$ point of the ESP luminosity
function (Zucca et al. 1997), also reproduce the observed exponential
cut-off at the bright end of the local B and K-band luminosity
functions (cf.  Fig.~\ref{fig:refLF}). This fact turns out to be of
importance when studying the clustering of these galaxies.

\begin{table}
\begin{center}
\caption{Variant models in $\tau$CDM cosmologies. The first column
gives the value of the parameter which is varied relative to the
reference model. The remaining columns give the values of $\Upsilon$
used to match a point in the luminosity function for each model and
the asymptotic bias of the galaxies estimated from the fitting formula
of Jing (1998), $b_{\mathrm analytic}$, and from our models,
$b_{\mathrm model}$.}
\label{tb:tCDMmods}
\begin{tabular}{lccc}
\hline \textbf{Model} & $\Upsilon$ & $\mathbf{b_{\mathrm{analytic}}}$ &
${\mathbf b}_{\mathrm{model}}$ \\ \hline Reference &1.23 & 1.27 & 1.65
$\pm$ 0.37 \\ $v_{\mathrm{hot}} = 350$ km/s &1.06 & 1.27 & 1.62 $\pm$
0.36 \\ $v_{\mathrm{hot}} = 200$ km/s &1.41 & 1.26 & 1.57 $\pm$ 0.29
\\ $\alpha _* = -0.25$ &1.20 & 1.27 & 1.62 $\pm$ 0.33 \\ $\alpha _* =
-1.50$ &1.33 & 1.27 & 1.63 $\pm$ 0.40 \\ $\epsilon _* = 0.01^{\dag}$
&0.98 & 1.27 & 1.64 $\pm$ 0.28 \\ $\epsilon _* = 0.04$ &1.52 & 1.29 &
1.72 $\pm$ 0.42 \\ $f_{\mathrm df} = 0.5^{\dag}$ &1.17 & 1.27 & 1.67
$\pm$ 0.44 \\ $f_{\mathrm df} = 0.03$ &1.23 & 1.25 & 1.59 $\pm$ 0.35
\\ IMF: Kennicutt (1993) &2.01 & 1.29 & 1.65 $\pm$ 0.38 \\
$r_{\mathrm{core}} = 0.2$ &1.22 & 1.26 & 1.60 $\pm$ 0.39 \\
$r_{\mathrm{core}} = 0.02$ &1.23 & 1.28 & 1.65 $\pm$ 0.36 \\ $\Omega
_{\mathrm{b}} = 0.10$ &1.82 & 1.28 & 1.59 $\pm$ 0.31 \\ $\Omega
_{\mathrm{b}} = 0.05$ & 0.55 & 1.26 & 1.71 $\pm$ 0.36 \\ $p = 0.02$
&1.20 & 1.29 & 1.69 $\pm$ 0.38 \\ Recooling &1.68 & 1.28 & 1.64 $\pm$
0.36 \\ No dust & 1.89 & 1.30 & 1.68 $\pm$ 0.36 \\ \hline
\end{tabular}
\end{center}
\begin{flushleft}
$^{\dag}$ The two models with the greatest deviation from the mean two-point correlation function.
\end{flushleft}
\end{table}

\begin{table}
\begin{center}
\caption{Variant models in $\Lambda$CDM cosmologies. The first column
gives the value of the parameter which is varied relative to the
reference model. The remaining columns give the values of $\Upsilon$
used to match a point in the luminosity function for each model and
the asymptotic bias of the galaxies estimated from the fitting formula
of Jing (1998), $b_{\mathrm analytic}$, and from our models,
$b_{\mathrm model}$.}
\label{tb:LCDMmods}
\begin{tabular}{lccc}
\hline \textbf{Model} & $\Upsilon$ & $\mathbf{b_{\mathrm{analytic}}}$
& ${\mathbf b}_{\mathrm{model}}$ \\ \hline Reference &1.63 & 1.07 &
1.01 $\pm$ 0.03 \\ $v_{\mathrm{hot}} = 200$ km/s &1.45 & 1.07 & 0.98
$\pm$ 0.01 \\ $v_{\mathrm{hot}} = 100$ km/s &1.53 & 1.06 & 0.97 $\pm$
0.02 \\ $\alpha _* = -0.25$ &1.63 & 1.08 & 0.98 $\pm$ 0.01 \\ $\alpha
_* = -1.50$ & 1.58 & 1.06 & 0.97 $\pm$ 0.01 \\ $\epsilon _* = 6.67
\times 10^{-3}$ &1.33 & 1.06 & 0.98 $\pm$ 0.01 \\ $\epsilon _* = 0.02$
&1.81 & 1.06 & 0.97 $\pm$ 0.01 \\ $f_{\mathrm df} = 5.0^{\dag}$ &1.29
& 0.93 & 0.88 $\pm$ 0.02 \\ $f_{\mathrm df} = 0.2^{\dag}$ &1.31 & 0.98
& 0.91 $\pm$ 0.01 \\ IMF: Salpeter (1955) & 0.90 & 1.01 & 1.00 $\pm$
0.02 \\ $r_{\mathrm{core}} = 0.2$ &1.63 & 1.07 & 0.98 $\pm$ 0.01 \\
$r_{\mathrm{core}} = 0.02$ &1.63 & 1.07 & 0.99 $\pm$ 0.01 \\ $\Omega
_{\mathrm{b}} = 0.04^{\dag}$ &2.04 & 1.13 & 1.02 $\pm$ 0.01 \\ $\Omega
_{\mathrm{b}} = 0.01^{\dag}$ &0.70 & 0.96 & 0.95 $\pm$ 0.02 \\ $p =
0.04^{\dag}$ &1.04 & 1.11 & 1.01 $\pm$ 0.01 \\ Recooling$^{\dag}$
&1.67 & 1.11 & 1.00 $\pm$ 0.02 \\ No dust &2.31 & 1.04 & 0.95 $\pm$
0.01 \\ \hline
\end{tabular}
\end{center}
\begin{flushleft}
$^{\dag}$ The six models with the greatest deviation from the mean two-point correlation function.
\end{flushleft}
\end{table}

\begin{figure*}
\begin{tabular}{cc}
\psfig{file=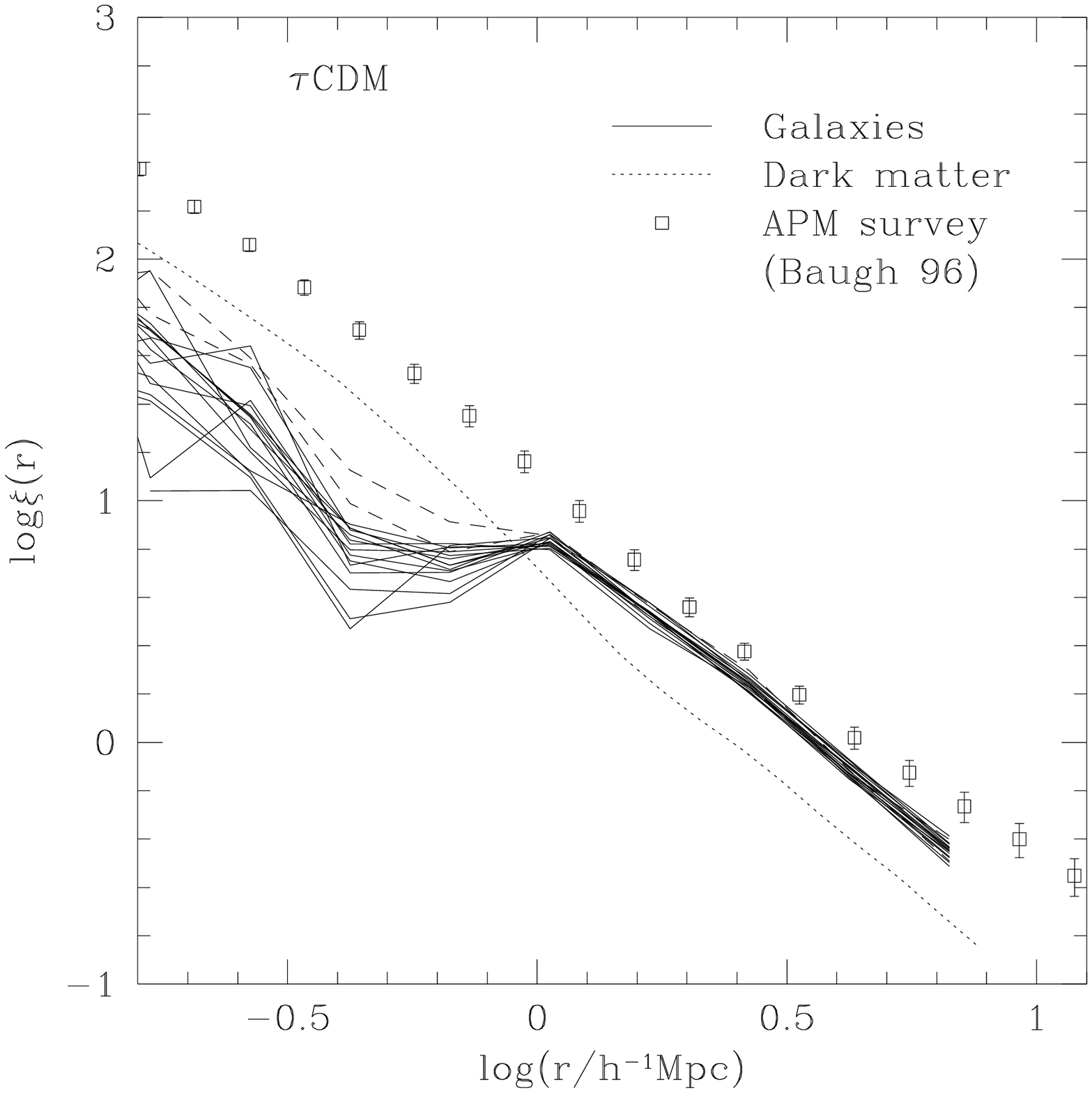,width=80mm} &
\psfig{file=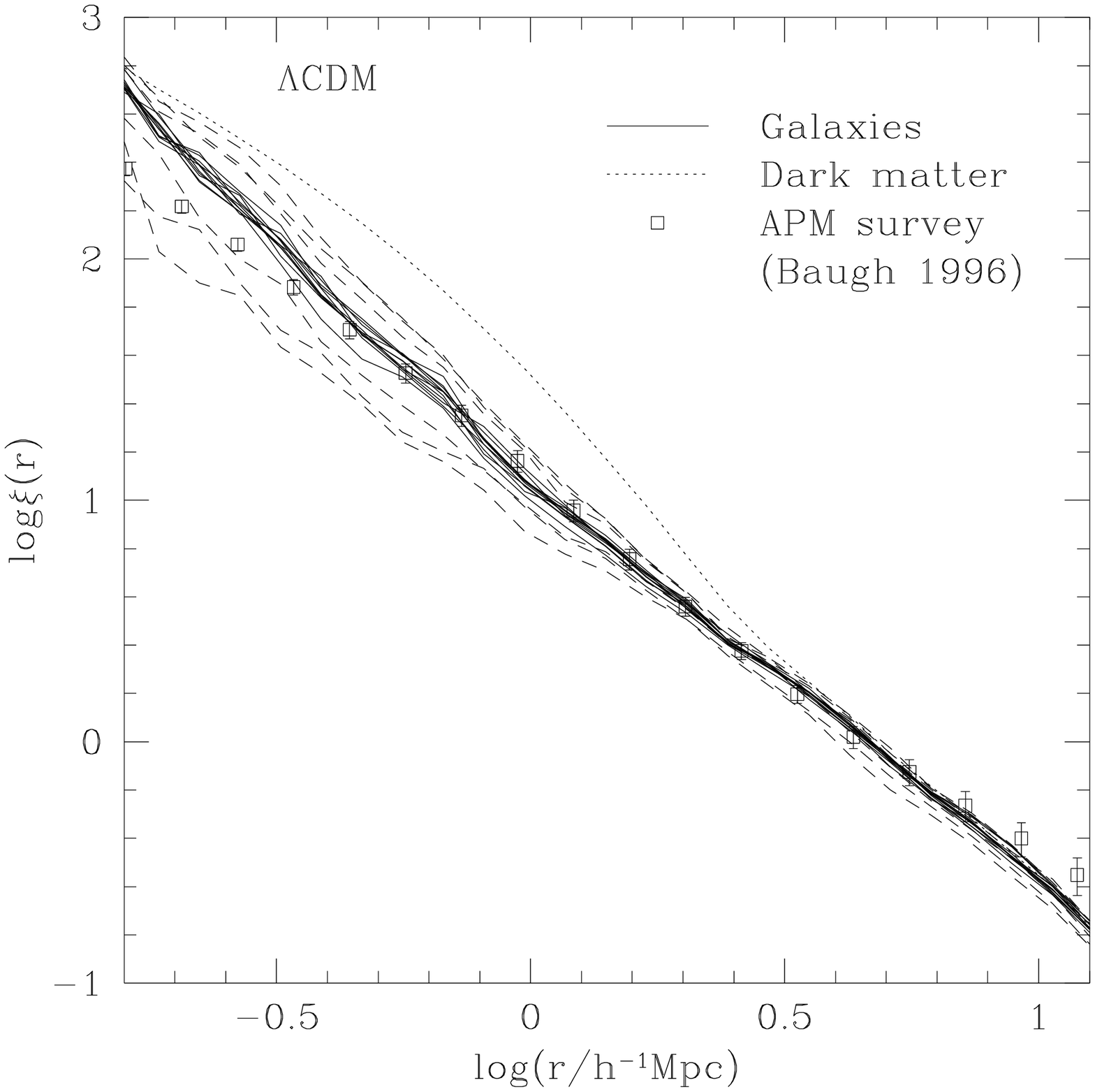,width=80mm}
\end{tabular}
\caption{Correlation functions in the $\tau$CDM (left hand panel) and
$\Lambda$CDM (right hand panel) cosmologies. All the models are constrained
to match the abundance of $L_*$ galaxies in the ESP B-band luminosity
function. In each plot the points with error bars show the observed APM
real-space correlation function of Baugh (1996), whilst the dotted line
shows the correlation function of the dark matter. The model galaxy
correlation functions are shown as solid lines except in the case of models
which deviate substantially from the average of all models which are shown
as dashed lines.}
\label{fig:xiLFnorm}
\end{figure*}

\begin{figure*}
\begin{tabular}{cc}
\psfig{file=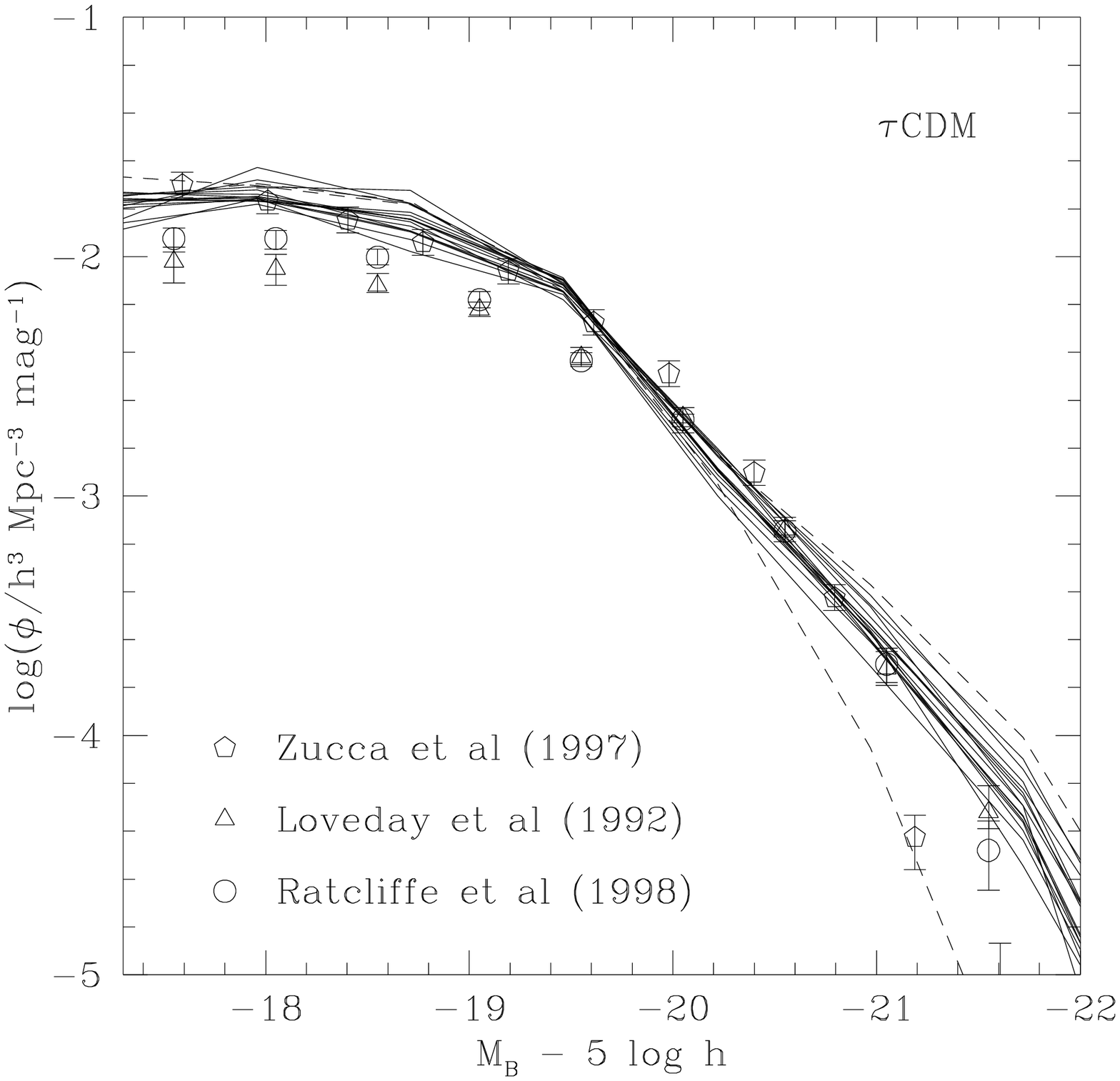,width=80mm} &
\psfig{file=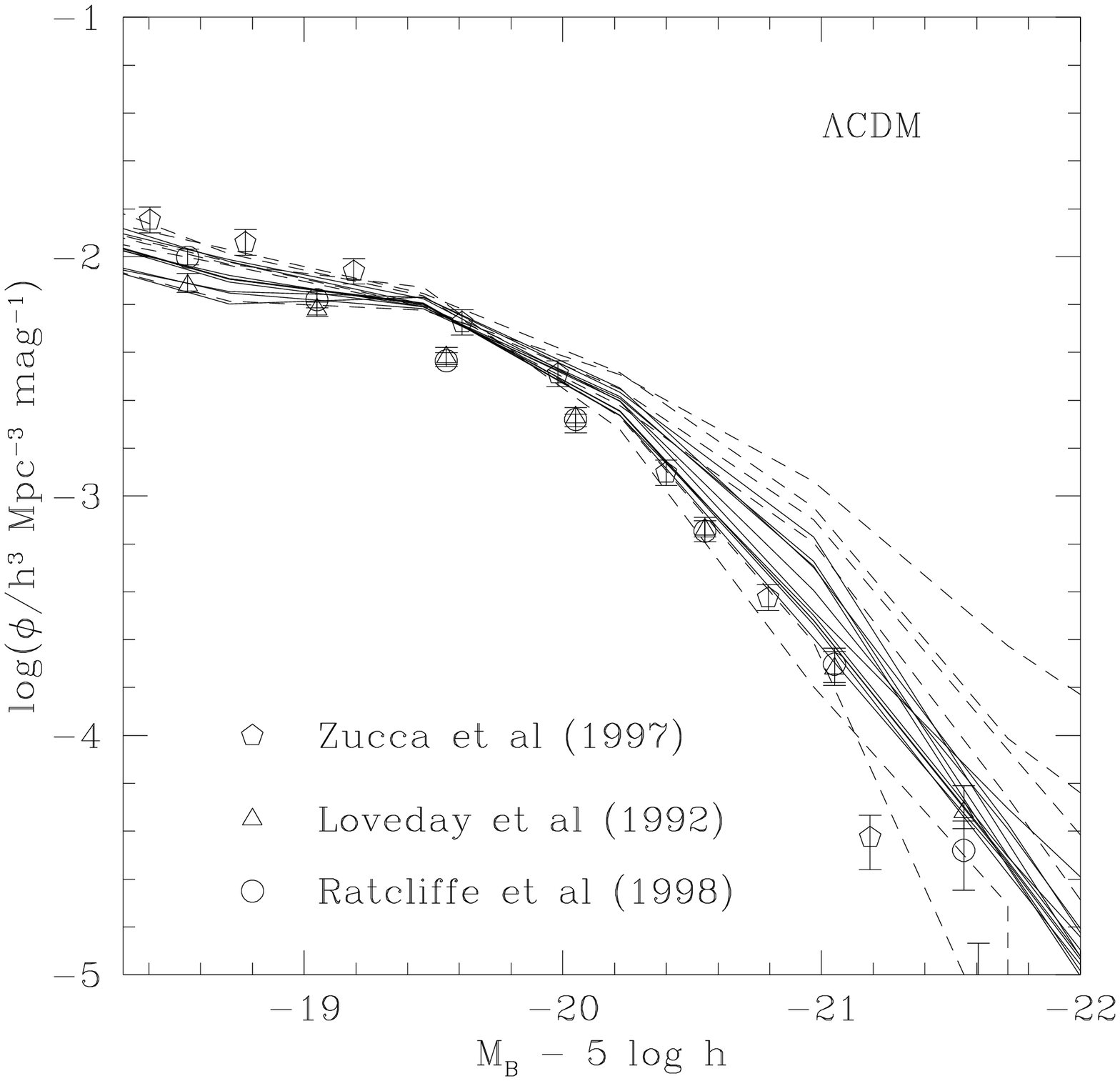,width=80mm}
\end{tabular}
\caption{B-band luminosity functions in the $\tau$CDM (left hand panel) and
$\Lambda$CDM (right hand panel) models. All models are constrained to match
the ESP luminosity function of Zucca et al. (1997) at $M_{\mathrm{B}} - 5
\log h = -19.56$. Symbols with error bars show a selection of observational
determinations of the luminosity functions, from the sources indicated in
the legend. The solid lines show results for our models, except that the
outliers identified in Fig.~\ref{fig:xiLFnorm} are shown as dashed
lines. Each luminosity function is plotted only to the completeness limit
of the simulations.}
\label{fig:LFLFnorm}
\end{figure*}

\begin{figure*}
\begin{tabular}{cc}
\psfig{file=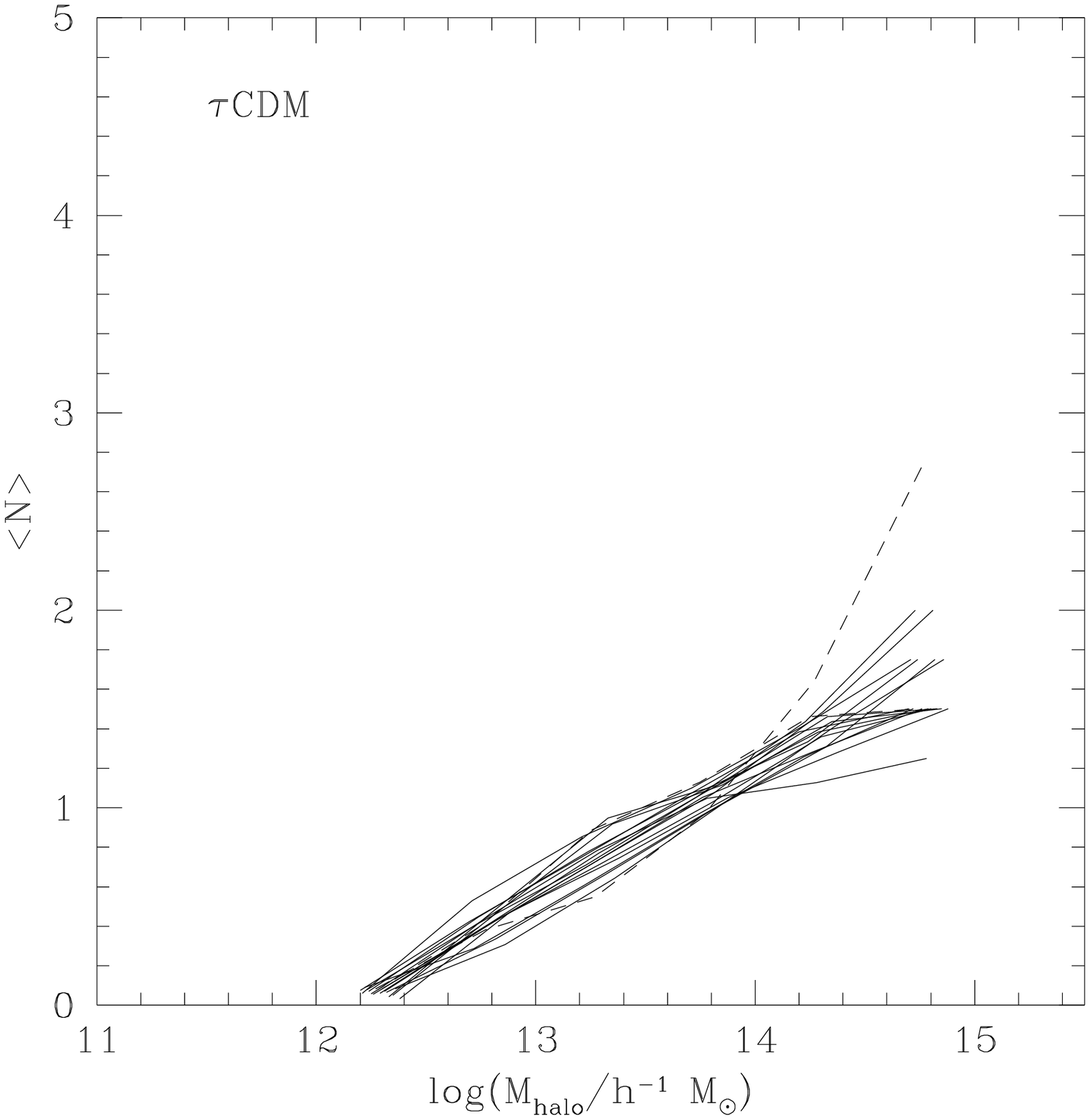,width=80mm} &
\psfig{file=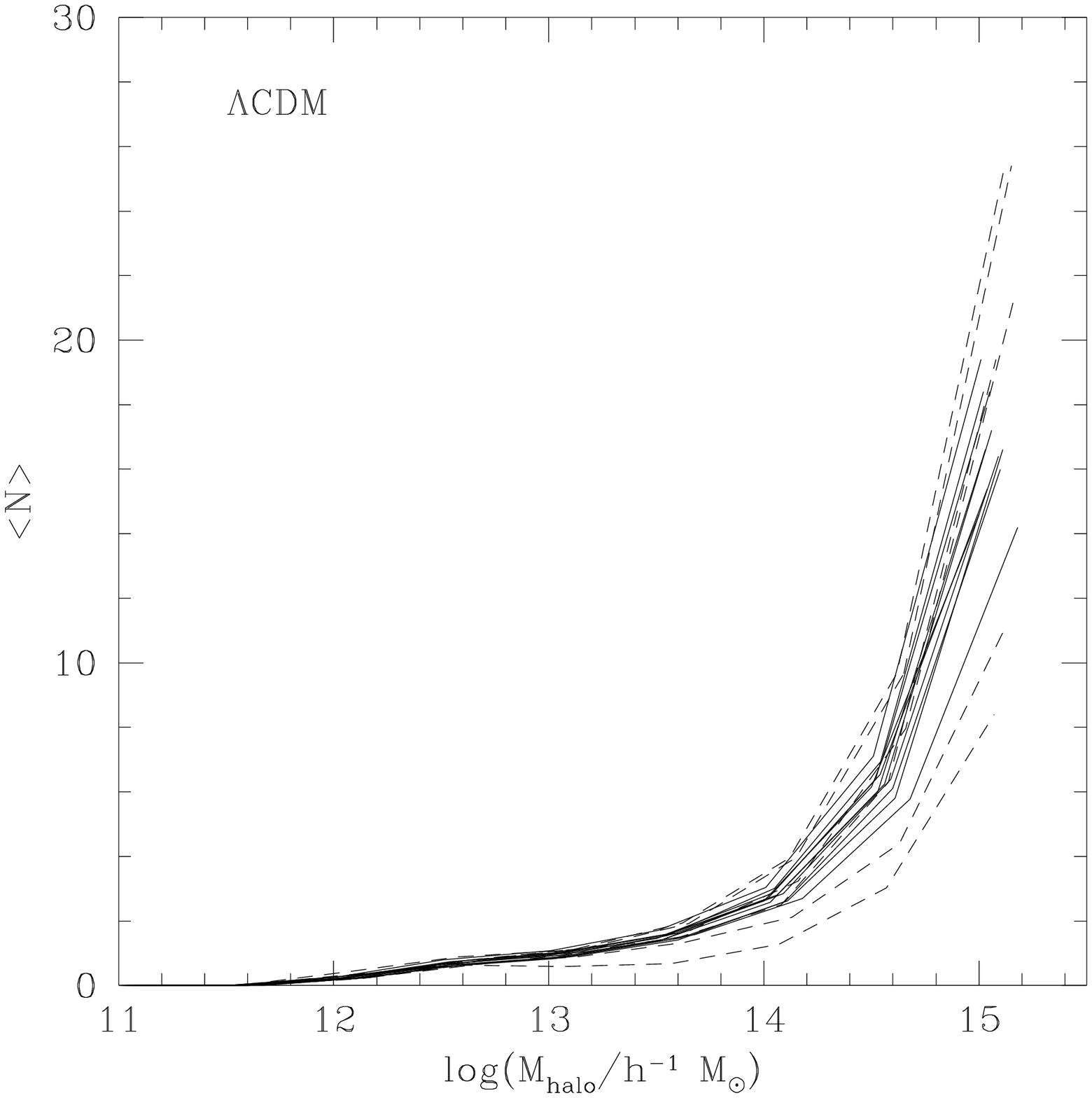,width=80mm}
\end{tabular}
\caption{The average number of galaxies brighter than $M_{\mathrm{B}} - 5
\log h = -19.5$ per halo as a function of halo mass. The panels refer to
the $\tau$CDM models (left) and the $\Lambda$CDM models (right). Note that
unoccupied halos are counted. Dashed lines indicate the outlier models
identified in Fig.~\ref{fig:xiLFnorm}. (Some of the lines lie on top of one
another because the corresponding changes in model parameters do not alter
the number of galaxies per halo.)}
\label{fig:NMLFnorm}
\end{figure*}

Tables \ref{tb:tCDMmods} \& \ref{tb:LCDMmods} list the variant models
that we have studied in the $\tau$CDM and $\Lambda$CDM cosmologies
respectively. The first column of each table lists those parameters
that have changed from the reference model (which is listed in the
first row of the tables). ``Recooling'' models allow enriched gas
reheated by supernovae to recool within a dark matter halo. All but
one model use the ``No recooling'' algorithm in which this gas is not
allowed to recool until its halo doubles in mass. In the ``No dust''
models we do not account for extinction by internal dust (see Cole et
al 1999). Also listed in the tables is the value of $\Upsilon$ for
each model and the average analytic asymptotic bias of the galaxies
calculated as follows:

\begin{equation}
b_{\mathrm analytic} = \sum ^N_{i=1} b(M_i)/N \equiv { \int _0^{\infty}
b(M) \bar{N} (M) n(M) {\mathrm d}M \over \int _0^{\infty} \bar{N} (M) n(M)
{\mathrm d}M},
\end{equation}

\noindent where $N$ is the number of galaxies in the catalogue, $M_i$ is
the mass of the halo hosting the $i^{\mathrm th}$ galaxy, $\bar{N} (M)$ is
the mean number of galaxies per halo of mass $M$, and $n(M)$ is the
dark matter halo mass function in the simulation. The function $b(M)$ is the
asymptotic bias of halos of mass $M$ which we estimate using Jing's (1998)
formula obtained from fitting the results of N-body simulations. This
formula tends to the analytic result of Mo \& White (1996) for masses much
greater than $M_*$. The final column gives the asymptotic bias estimated
directly from our models, on scales where $\xi _{\mathrm matter} (r) < 1$
($\approx 2.5$ and $5.0 h^{-1}$ Mpc in the $\tau$CDM and $\Lambda$CDM
cosmologies respectively), as described by Jing (1998). Note that the
analytic biases are consistently lower than those measured in our $\tau$CDM
models. The halos in the $\tau$CDM GIF simulation show a similar
disagreement with the fitting formula of Jing (1998) which was tested on
SCDM, OCDM and $\Lambda$CDM cosmologies only. The models marked by a dagger
are those showing large deviations from the mean clustering amplitude of
all models (six and two such models are identified in the $\Lambda$CDM and
$\tau$CDM cosmologies respectively).

In the remainder of this section we show the correlation functions
obtained from these variant models and discuss how the form of the
correlation function is related to other properties of the galaxy
population. The correlation functions are displayed in
Fig.~\ref{fig:xiLFnorm}. All cases show antibias on small scales and a
constant bias on large scales. Most of the models in both cosmologies
have similar correlation functions but the scatter is somewhat greater
in the $\tau$CDM case than in the $\Lambda$CDM case. The models that
deviate most from the average are shown as dashed lines in
Fig.~\ref{fig:xiLFnorm} and also in all other plots in this
section. In the $\Lambda$CDM cosmology, where the reference model is
well fit by a power-law correlation function, the deviant models have
slopes which are somewhat different from the other models.

The luminosity functions in most of the $\Lambda$CDM models, plotted
as solid lines, in Fig.~\ref{fig:LFLFnorm}, are quite similar. (They
are all forced to go through the same point at $M_{\mathrm B} - 5 \log
h = -19.6$.)  The ones that deviate the most are those plotted as
dashed lines, that is, those that were identified in
Fig.~\ref{fig:xiLFnorm} as giving the most discrepant correlation
functions. Thus, we see that the main factor that determines the
sensitivity of the correlation function to model parameters is the
ability of the model to reproduce the exponential cut-off observed in
the luminosity function. Models that achieve this all give similar
galaxy correlation functions. A similar conclusion applies in the
$\tau$CDM case, although here the distinction is less clearcut due to
our noisy estimates of the correlation function on small scales.

We have also considered two models that have different dark matter
power spectra. These make use of the GIF SCDM and OCDM simulations
(described in \S\ref{sec:desc}) and, apart from the values of the
cosmological parameters, they have the same parameter values as our
reference $\tau$CDM and $\Lambda$CDM models respectively. Whilst the
OCDM model shows very little difference from the $\Lambda$CDM model,
the SCDM model has a significantly different clustering amplitude than
our reference $\tau$CDM model (approximately 40\% lower on scales
larger than 1 Mpc). Despite this its luminosity function is in fairly
close agreement with that of the $\tau$CDM model at the bright
end. This model therefore demonstrates that models with the same
luminosity function only produce the same correlation function if they
have the same underlying dark matter distribution.

As described in \S\ref{sec:2pt} the distribution of the number of
galaxies per halo as a function of halo mass is very important in
determining the behaviour of the correlation function. On small
scales, the full distribution determines the amplitude and slope of
the correlation function, whilst on large scales the number of
galaxies per halo determines the asymptotic bias of the galaxy
distribution by selecting the range of host halo masses that dominates
the correlation function.  Fig.~\ref{fig:NMLFnorm} shows the number of
galaxies per halo as a function of halo mass in our models. It is
apparent, particularly for $\Lambda$CDM, that the models identified as
outliers in the correlation function plot (Fig.~\ref{fig:xiLFnorm})
are also the ones that deviate the most from the reference models in
these plots as well.

\begin{figure*}
\begin{center}
\begin{tabular}{cc}
\psfig{file=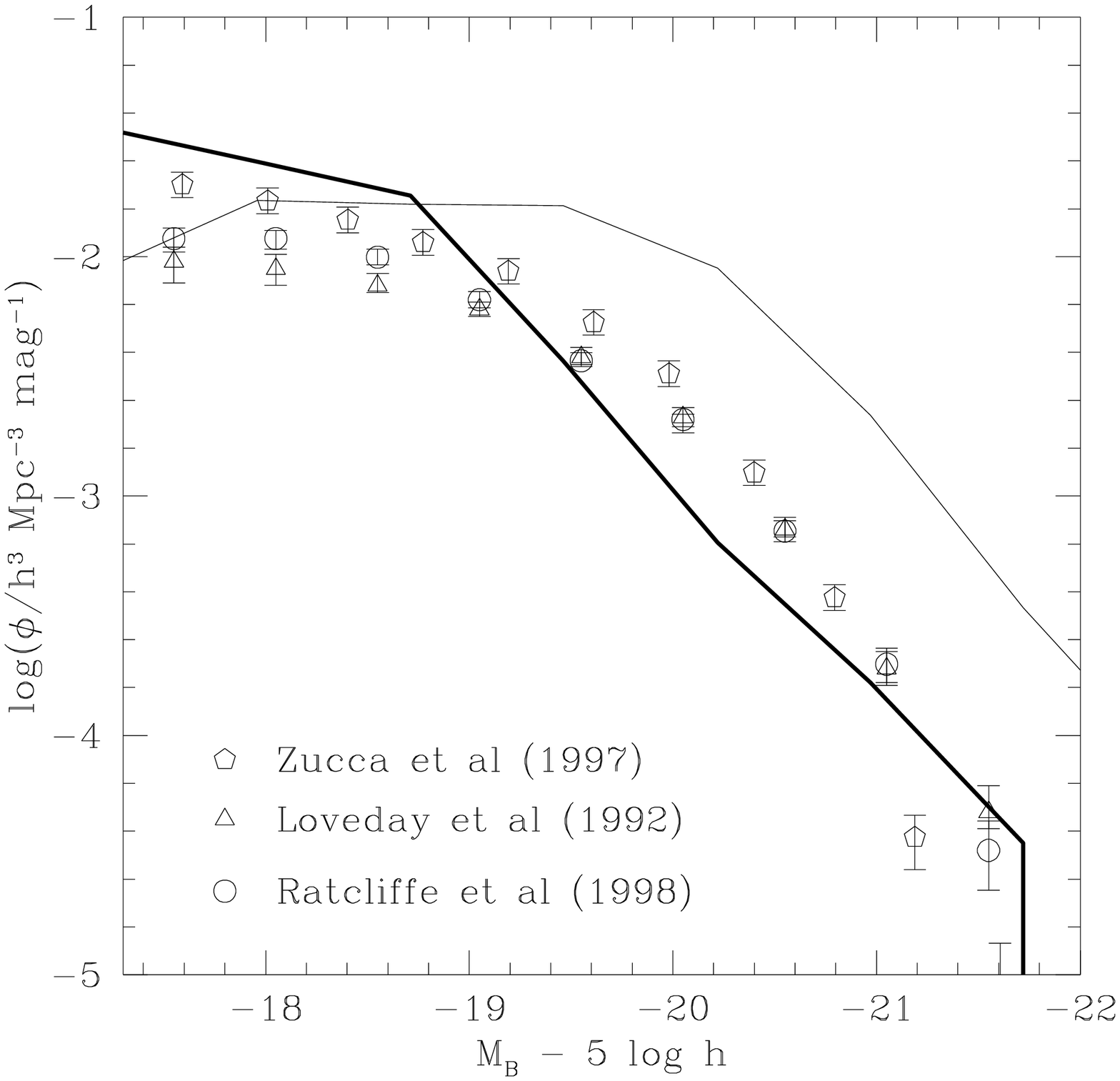,width=80mm} &
\psfig{file=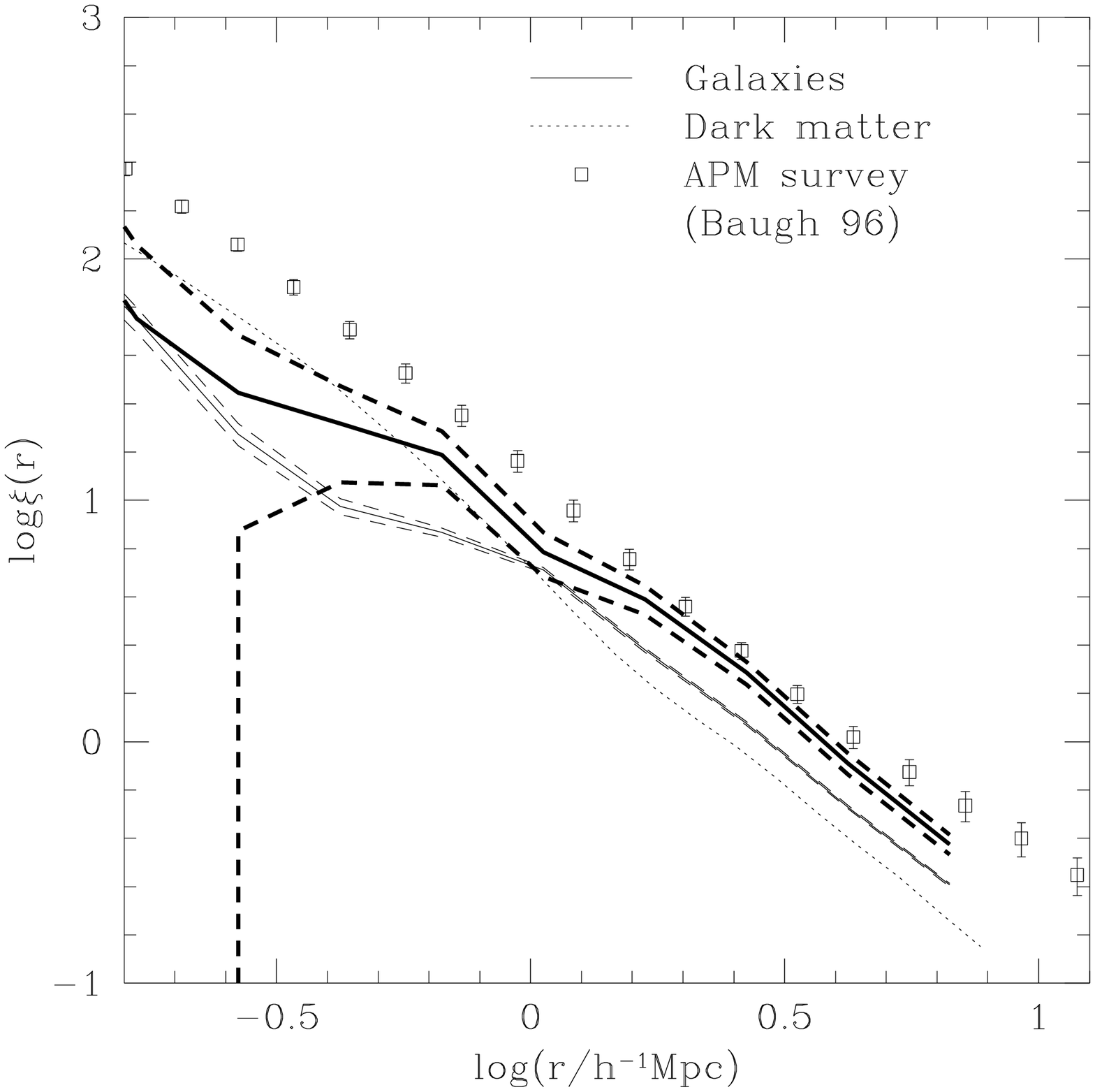,width=80mm}
\end{tabular}
\caption{The B-band luminosity function (left) and galaxy correlation
function (right) for two $\tau$CDM models constrained to match the I-band
Tully-Fisher relation. The thin line in both panels corresponds to our
reference model (but with the value of $\Upsilon$ required by the
Tully-Fisher relation) and the thick lines corresponds to a model with very
weak feedback. The luminosity function is shown only to the completeness
limit of this model. In the right hand panel the symbols with error bars
show Baugh's (1996) APM correlation function, the dashed lines the
Poissonian errors on the correlation functions, and the the dotted line the
dark matter correlation function.}
\label{fig:nofeedb}
\end{center}
\end{figure*}

\subsection{Models constrained by the Tully-Fisher relation}
\label{sec:TFnorm}

We have shown that matching the local galaxy luminosity function --
our preferred method for constraining the parameters of our
semi-analytic model -- leads to model predictions for galaxy
clustering that are robust to reasonable changes in these
parameters. Kauffmann et al. (1999a) adopted a different philosophy:
they chose to constrain their models by matching the I-band
Tully-Fisher relation, rather than the luminosity function. We explore
the effect of this choice by constraining our own models in a similar
way. Specifically, we require the median magnitude of central spiral
galaxies with halo circular velocities in the range 215.0 to 225.0 km
s$^{-1}$ to be $M_{\mathrm{I}} - 5 \log h = -22.0$ (where $M_{\rm I}$
is the dust-extincted I-band magnitude of each galaxy corrected to the
face-on value). This can be achieved by a suitable choice of the
luminosity normalisation parameter, $\Upsilon$, and leads to a model
Tully-Fisher relation that agrees well with data from Mathewson, Ford
\& Buchhorn (1992).

Since our original $\Lambda$CDM models (that is, the reference model
and its variants) already agreed quite well with the Tully-Fisher
relation, (see Fig.~\ref{fig:refTF}), this different choice of
constraint has only a minor effect on the correlation function. The
only noticeable change is an increase in the scatter of the asymptotic
bias in the variant models. In the $\tau$CDM models, on the other
hand, the new constraint has an important effect because the original
models that agreed well with the luminosity function, missed the
Tully-Fisher relation by about 1 magnitude. Forcing a fit to the
Tully-Fisher relation destroys the good agreement of the reference
model with the luminosity function, as may be seen in
Fig.~\ref{fig:nofeedb}. This figure also shows a $\tau$CDM model in
which we have attempted to obtain a better luminosity function by
dramatically reducing the amount of supernovae feedback into the
interstellar gas.

Fig.~\ref{fig:nofeedb} shows that our two $\tau$CDM models that match the
Tully-Fisher relation have different correlation functions. In other words,
this exercise demonstrates that when models are constrained in this way, the
resulting correlation functions are rather sensitive to the choice of model
parameters. This explains why Kauffmann et al. (1999a) concluded that their
clustering predictions depended strongly on the way they parametrised star
formation, feedback and the fate of reheated gas in their model. By
contrast, we have found that our predictions for the correlation function
are robust to changes in model parameters, {\it so long as the models match
the bright end of the galaxy luminosity function}.

\section{Discussion and Conclusions}
\label{sec:discuss}

In this study, we have considered some of the physical and statistical
processes that determine the distribution of galaxies and its relation to
the distribution of mass in cold dark matter universes. The approach that
we have adopted exploits two of the most successful techniques currently
used in theoretical cosmological studies: N-body simulations to follow the
clustering evolution of dark matter and semi-analytic modelling to follow
the physics of galaxy formation. Our main conclusion is that the efficiency
of galaxy formation depends in a non-trivial fashion on the mass of the
host dark matter halo and, as a result, galaxies, in general, have a
markedly different distribution from the mass. This result had been
anticipated in early cosmological studies (eg. Frenk, White \& Davis 1983,
Davis et al. 1985, Bardeen et al. 1986), but it is only with the
development of techniques such as semi-analytic modelling that realistic
calculations have become possible.

The statistics of the spatial distribution of galaxies reflect the
interplay between processes that determine the location where dark
matter halos form and the manner in which halos are ``lit up" by
galaxy formation. If the resulting mass-to-light ratio of halos were
independent of halo mass, then the distribution of galaxies would be
related in a simple manner to the distribution of dark matter in
halos. In current theories of galaxy formation, however, the
mass-to-light ratio has a complicated dependence on halo mass. On
small mass scales, galaxy formation is inhibited by the reheating of
cooled gas through feedback processes, whereas in large mass halos it
is inhibited by the long cooling times of hot gas. As a result, the
mass-to-light ratio has a deep minimum at the halo mass, $\sim 10^{12}
M_{\odot}$, associated with $L_*$ galaxies, where galaxy formation is
most efficient. Although our calculations assume a specific
model of galaxy formation, the dependence of mass-to-light ratio on
halo mass displayed in Fig.~\ref{fig:MLLF} is likely to be generic to
this type of cosmological model. The consequence of such a complex
behaviour is a scale dependent bias in the distribution of galaxies
relative to the distribution of mass.

On scales larger than the typical size of the halos that harbour
bright galaxies, the bias in the galaxy distribution is related in a
simple way to the bias in the distribution of massive halos. In our
$\Omega _0 =1$ $\tau$CDM model, galaxies end up positively biased on
large scales, but in our flat, $\Omega _0 =0.3$ $\Lambda$CDM model,
they end up essentially unbiased. On small scales, the situation is
more complicated and the correlation function depends on effects such
as the spatial exclusion of dark matter halos, dynamical friction, and
the number of galaxies per halo.  In particular, our simulations show
how the statistics of the halo occupation probability influence the
amplitude of the galaxy correlation function on sub-megaparsec
scales. In our models, the occupation of halos by galaxies is not a
Poisson process. Since the amount of gas available for star formation
is limited, the mean number of pairs per halo is less than that of a
Poisson distribution with the same mean. This property plays an
important role in determining the amplitude of small scale
correlations.

Remarkably, the correlation function of galaxies in our $\Lambda$CDM
model closely approximates a power-law over nearly four orders of
magnitude in amplitude. This is in spite of the fact that the
correlation function of the underlying mass distribution is not a
power-law, but has two inflection points in the relevant range of
scales. Somehow, the various effects just discussed conspire to
compensate for these features in the mass distribution. In particular,
on scales smaller than $\sim 3 h^{-1}$ Mpc, the galaxy distribution in
the $\Lambda$CDM model is {\it antibiased} relative to the mass
distribution. The apparently scale-free nature of the galaxy
correlation function in this model seems to be largely a coincidence
(although whether this is also true of the real universe remains to be
seen). Our $\tau$CDM model which has similar physics
although a different initial mass fluctuation spectrum, does not end
up with a power-law galaxy correlation function.

Col\'{\i}n et al. (1999) have carried out a very high resolution
N-body simulations (of a $\Lambda$CDM cosmological model similar to
ours) which resolves some substructure within dark matter halos. The
correlation function of these sub-halos is remarkably similar to the
correlation function of the galaxies in our $\Lambda$CDM reference
model. In a sense, the merger trees of our semi-analytic models keep
track of sub-halos within dark matter halos since they follow the
galaxies that form within them. (Unlike the simulations, however, the
semi-analytic model does not follow the spatial distribution of
sub-halos.)  Col\'{\i}n et al. select sub-halos by circular velocity,
whereas we select galaxies by luminosity. Since, in our models, the
luminosity of a galaxy is correlated with the circular velocity of the
halo in which it formed, there is some correspondence between the type
of objects studied by Col\'{\i}n et al. and us. However, the
connection could be complicated by effects such as tidal disruption or
stripping of halos within halos but which are not included in our
model. Nevertheless, the abundance of sub-halos considered by
Col\'{\i}n et al is similar to the abundance of galaxies in our
$\Lambda$CDM model and this might account for the similarity between
the two correlation functions.

Another noteworthy outcome of our simulations is the close match of
the galaxy correlation function in our $\Lambda$CDM model to the
observed galaxy correlation function, itself also a power-law over a
large range of scales (Groth \& Peebles 1977; Baugh 1996). This match
is particularly interesting because the parameters that specify our
semi-analytic galaxy formation model were fixed beforehand by
considerations that are completely separate from galaxy clustering
(see Cole et al. 1999). Our procedure for fixing these parameters
places special emphasis on obtaining a good match to the observed
galaxy luminosity function (c.f. Fig.~\ref{fig:refLF}), but makes no
reference whatsoever to the spatial distribution of the galaxies. 

To summarize, the combination of high resolution N-body simulations with
semi-analytic modelling of galaxy formation provides a useful means for
understanding how the process of galaxy formation interacts with the
process of cosmological gravitational evolution to determine the clustering
pattern of galaxies. In general, we expect galaxies to be clustered
somewhat differently from the dark matter, and the relation between the two
can be quite complex. A flat CDM model with $\Omega_0=0.3$ gives an
acceptable match to the observed galaxy correlation function over about
four orders of magnitude in amplitude (as does an open model with the same
value of $\Omega_0$.) The $\Lambda$CDM model is also in reasonable
agreement with a number of other known properties of the galaxy
distribution.

\section*{Acknowledgements}

AJB, SMC and CSF acknowledge receipt of a PPARC Studentship, Advanced
Fellowship and Senior Fellowship respectively. CSF also acknowledges a
Leverhume Fellowship. CGL was supported by the Danish National Research
Foundation through its establishment of the Theoretical Astrophysics
Center. This work was supported in part by a PPARC rolling grant, by a
computer equipment grant from Durham University and by the European
Community's TMR Network for Galaxy Formation and Evolution. We acknowledge
the Virgo Consortium and the GIF collaboration for making available the GIF
simulations for this study, and the anonymous referee for insightful and 
helpful comments.

\end{document}